\documentclass[aps,10pt,twocolumn,showpacs,citeautoscript,amsmath,amssymb,floatfix,superscriptaddress]{revtex4-1}

\usepackage{graphicx}
\usepackage{dcolumn}
\usepackage{bm}
\usepackage{epsfig,hyperref,amsthm}
\usepackage{mathtools}
\usepackage{color}
\usepackage{natmove}
\usepackage{colordvi}
\usepackage[dvipsnames]{xcolor}
\usepackage{relsize}
\usepackage{accents}
\usepackage{wasysym}  
\usepackage[caption=false]{subfig}

\hypersetup{
	colorlinks=true,
	linkcolor=blue,
	citecolor=blue,
	urlcolor=blue
}

\newcommand{\beq}{\begin{equation}}
\newcommand{\eeq}{\end{equation}}
\newcommand{\beqar}{\begin{eqnarray}}
\newcommand{\eeqar}{\end{eqnarray}}
\newcommand{\bea}{\begin{eqnarray}}
\newcommand{\eea}{\end{eqnarray}}
\newcommand{\bcen}{\begin{center}}
\newcommand{\ecen}{\end{center}}

\newcommand{\bra}[1]{\left< #1 \right|}
\newcommand{\ket}[1]{\left| #1 \right>}
\newcommand{\ketbra}[2]{\left| #1 \right> \left< #2 \right|}

\newcommand{\ave}[1]{\left< #1 \right>}
\newcommand{\dg}{^\dagger}

\newcommand{\ad}{a^{\dagger}}
\newcommand{\bd}{b^{\dagger}}
\newcommand{\p}{_{+}}
\newcommand{\m}{_-}
\newcommand{\tr}[2]{\mathrm{Tr}_{#1}\left[ #2 \right]}

\begin{document}

\title{Response theory for nonequilibrium steady-states of open quantum systems}

\author{Amikam Levy}
\email{amikamlevy@gmail.com}
\affiliation{Department of Chemistry, Bar-Ilan University, Ramat-Gan 52900, Israel}

\author{Eran Rabani}
\email{eran.rabani@berkeley.edu}
\affiliation{Department of Chemistry,  University of California, Berkeley, California 94720, United States}
\affiliation{The Raymond and Beverly Sackler Center for Computational Molecular and Materials Science, Tel Aviv University, Tel Aviv, Israel 69978
}
\affiliation{Materials Sciences Division, Lawrence Berkeley National Laboratory, Berkeley, California 94720, United States}

\author{David T. Limmer}
\email{dlimmer@berkeley.edu}
\affiliation{Department of Chemistry,  University of California, Berkeley, California 94720, United States}
\affiliation{Materials Sciences Division, Lawrence Berkeley National Laboratory, Berkeley, California 94720, United States}
\affiliation{Chemical Sciences Division, Lawrence Berkeley National Laboratory, Berkeley, California 94720, United States}
\affiliation{Kavli Energy NanoScience Institute, Berkeley, California 94720, United States}

\begin{abstract}
 We introduce a response theory for open quantum systems within nonequilibrium steady-states subject to a Hamiltonian perturbation. Working in the weak system-bath coupling regime, our results are derived within the Lindblad-Gorini-Kossakowski-Sudarshan formalism.
We  find that the response of the system to a small perturbation is not simply related to a correlation function within the system, unlike traditional linear response theory in closed systems or expectations from the fluctuation-dissipation theorem. In limiting cases, when the perturbation is small relative to the coupling to the surroundings or when it does not lead to a change of the eigenstructure of the system, a perturbative expansion exists where the response function is related to a sum of a system correlation functions and additional forces induced by the surroundings. Away from these limiting regimes however, the secular approximation results in a singular response that cannot be captured within the traditional approach, but can be described by reverting to a microscopic Hamiltonian description. These  findings are illustrated by explicit calculations in coupled qubits and anharmonic oscillators in contact with bosonic baths at different temperatures.
%
\end{abstract}

\maketitle

\section{Introduction}

Response theory provides a means of characterizing the dynamical behavior of systems subject to small perturbations. For physical systems obeying detailed balance, the fluctuation-dissipation theorem relates a system's response to the underlying microscopic correlations~\cite{zwanzig2001nonequilibrium}. Such fluctuation-response relations form the basis of molecular spectroscopy~\cite{kubo1962stochastic,klauder1962spectral,mukamel1995principles} and are encoded in properties like the conductivities of transport devices and sensitivities of sensors~\cite{kubo2012statistical,hone1999thermal,vadlamani2019tunnel}. For open quantum systems, connections between response and molecular degrees of freedom are generally unknown in instances away from thermal equilibrium. Given the ubiquitous presence of open quantum systems, it is desirable to uncover such relations. Here we work within the Lindblad-Gorini-Kossakowski-Sudarshan (LGKS) description of an open quantum system and develop a response theory for dynamic perturbations~\cite{Lindblad74,gorini276,gorini76}.
We find that it is not typically possible to write the response function as a simple correlation function within the system, due to the weak coupling approximation and the reduced description of the LGKS formalism.
Nevertheless, we are able to find practical general routes to response functions and identify special cases where Dyson-like expansions of the propagator still exist, resulting in generalized fluctuation-dissipation relations. 

In the linear response regime for systems obeying detailed balance, response theory has revealed the fluctuation-dissipation theorems that establish
a relation between a system's response to a small perturbation and a time correlation function of certain observables evaluated at equilibrium.  For a closed quantum system that follows a unitary dynamics, the linear response function of an observable $A$ to a weak perturbation $\tilde{\delta} (t)V(t)$ is given by the well known Kubo formula \citep{kubo57} 
\beq
\label{eq:Kubo}
\phi_A(\tau-t) \equiv \frac{d \ave{A(\tau)}}{d \tilde{\delta} (t)} =\frac{i}{\hbar} \ave{[V(t),A(\tau)]}
\eeq
where $V$ is the perturbation operator added to the original Hamiltonian and $\tilde{\delta}(t)$ is a small time dependent parameter, and $\hbar$ is Planck's constant. 
 The brackets, $\ave{ \dots }$, denote average over the stationary distribution of the original Hamiltonian. Kubo's result has a classical analogue uncovered originally by Callen and Welton,\cite{callen1951irreversibility} and  has been extended to stochastic nonequilibrium steady states (NESS's)\citep{agarwal1972fluctuation,bochkov1981nonlinear,hanggi1982stochastic,marconi2008fluctuation} and nonlinear response~\cite{andrieux2006fluctuation,basu2015frenetic,gao2019nonlinear,lesnicki2020field}.  Recent work has also established relationships between these results and the entropy production within stochastic thermodynamics~\citep{seifert2010fluctuation}.

To extend these results to quantum systems that continuously interact with their environment, either the environment needs to be consider explicitly or a suitable reduced description found. While the former are 
well suited to accounting for instances of strong system-bath coupling~\cite{bruch2016quantum,ochoa2018quantum,bruch2018landauer,dou2018universal,dou2020universal}, describing nonequilibrium steady-states in such formalisms are cumbersome, as the composite system plus environment represents a closed system incapable of dissipating energy. If the couplings between the system and environment can be assumed to be weak, a reduced description using the LGKS master equation is appropriate, and a linear response theory can be formulated in the reduced Liouville space~\cite{chetrite2012quantum,venuti2016dynamical,mehboudi2018fluctuation,vznidarivc2019nonequilibrium,konopik2019quantum,albert2016geometry,mehboudi2019linear}. However, studies along these lines have thus far been largely phenomenological.

\begin{figure}
	\includegraphics[width=0.85\linewidth]{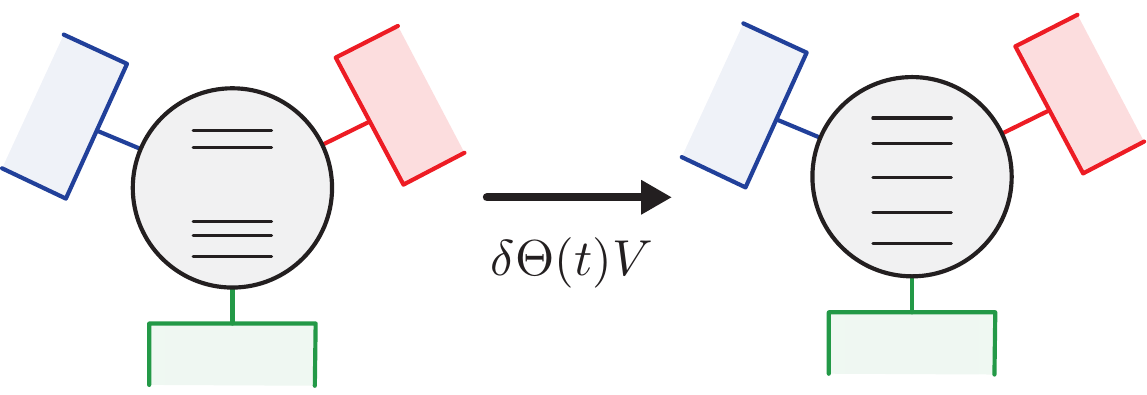}
	\caption{Illustration of the perturbations with nonequilibrium steady-states of open quantum systems considered.}
\label{fig:schematic}
\end{figure}

In this work, we introduce a response theory for open quantum systems within the LGKS formalism that is consistent with an underlying Hamiltonian description of the full system, sub-system plus environment.   
 In particular, we study the response of an open quantum system at equilibrium or NESS to a small \textit{Hamiltonian perturbation}. The initial NESS of the system corresponds to a steady-state of a system interacting with multiple baths. The system, which remains in contact with the baths at all times, is then subject to a stationary Hermitian perturbation that drives it away from its initial steady-state, as illustrated in Fig.~\ref{fig:schematic}. 

We find that even a small  Hamiltonian perturbation with respect to the system Hamiltonian may result in nonlinear response, and that unlike the case for closed quantum systems or classical systems, the corresponding response function is not simply related to a correlation function of the system. This occurs when the perturbation is considered small with respect to the system Hamiltonian, but not necessarily small with respect to the system-bath Hamiltonian. In such cases, additional timescales emerge within the system whose implications to the Markovian and secular approximations in the LGKS formalism must be considered. 

In cases where the perturbation is small with respect to the system-bath Hamiltonian, or when the perturbation influences  only the eigenvectors of the system Hamiltonian, a linear response relation can be established.  However, we find that the linear response function is subject to corrections over the closed system result that can be traced to noncommutative effects between the perturbation and the system-bath coupling Hamiltonians. Only with the inclusion of these additional terms is the response thermodynamically consistent.

The observation that the perturbation affects the system-bath couplings and leads to corrections in the response function is closely linked to the local and global approaches to deriving the LGKS master equation \cite{levy14b}.
In the local approach, 
the dynamics of a quantum system composed of several subsystems, in which each is coupled to its own surroundings, is described by a local dissipator that ignores the couplings between the subsystems.
The global approach takes the inter-couplings between the subsystems into consideration when deriving the dissipative part. 
The validity of the different approaches has been studied extensively in recent years and is an ongoing inquiry \cite{levy14b,barra2015thermodynamic,trushechkin2016perturbative,purkayastha2016out,gonzalez2017,hofer17,de2018reconciliation,naseem2018thermodynamic,farina2019open,cattaneo2019local,benatti2020bath,giorgi2020microscopic,elouard2020thermodynamics}.
This study offers a systematic method to calculate the response functions of open quantum systems, and, in doing so, bridges the two approaches. In particular, using a perturbative treatment, we extend the validity of the local approach without the need of a full global treatment, which may become complicated and, in most cases, not analytically feasible. 

The manuscript is laid out in the following way. First in Sec.~\ref{sec:perturb}, we briefly review the assumptions invoked in the LGKS framework, pay particular attention to the coarse-graining required to satisfy the Markovian approximation. Then in Sec.~\ref{sec:RespT}, we develop a response theory for impulsive system perturbations. This is specialized to instances where only the eigenvectors change in Sec.~\ref{sec:eigenvector}, with an explicit example of such a theory for two anharmonic oscillators are coupled to two heat baths considered in Sec.~\ref{sec:example1}. 
A response theory  specialized to instances where only the eigenvalues change is shown in Sec.~\ref{sec:eigenvalues}, with an explicit example of such a theory for two qubits coupled to two heat baths considered in Sec.~\ref{sec:example2}. 
The response function of thermodynamic functions that are state dependent is discussed in Sec.~\ref{sec:heatf}. 
Finally, in Sec.~\ref{sec:outlook} we summarize and discuss the main results.

\section{Framework for the LGKS master equation} 
\label{sec:perturb}
We consider a system-bath model with total Hamiltonian $H$ of the form
\beqar
H = H_{0} + \delta \theta(t)V + H_{sb}  + H_{b}
\eeqar
where the quantum system of interest is described by the unperturbed Hamiltonian, 
\beqar
H_{0}=\sum_n E^{(0)}_n \ket{\psi_{n}^{(0)}} \bra{\psi_{n}^{(0)}}
\eeqar
in its eigenbasis. 
While the baths may exist in the thermodynamic limit, we consider explicitly system Hilbert spaces that are discrete.
We focus our analysis on step perturbations $\tilde{\delta}(t)=\delta \theta(t)$, and $\delta$ is a unitless parameter that sets the scale, $\theta(t)$ is the Heaviside function turned on at $t=0$, and $V$ an arbitrary operator in the system Hilbert space. The system is weakly coupled to two or more baths held in distinct thermal states that can generate flows of mass and energy through the system described by
\beqar
H_{b}&=&\sum_\alpha \sum_j  \hbar \omega_{\alpha,j} r_{\alpha,j}^\dagger r_{\alpha,j}
\eeqar
where $r_{\alpha,j}^\dagger r_{\alpha,j}$ is the number operator for the $j$th mode of the $\alpha$th bath, which is assumed have a continuous spectrum and obey bosonic statistics. We will assume that the interaction with a bath is bilinear such that the interaction Hamiltonian is 
\beqar
H_{sb}=\sum_\alpha \lambda_\alpha S_\alpha R_\alpha
\eeqar
where for the $\alpha$th bath an operator in the system, $S_\alpha$, is coupled to a bath operator, $R_\alpha$, with strength $\lambda_{\alpha}$.

For a system interacting weakly with each bath in the absence of the perturbation $\delta=0$, assuming the baths are thermal and initially uncorrelated with each other, the reduced state of the system reads $\rho(t)\approx~\prod_\alpha \exp[\lambda_{\alpha}^2 K_\alpha (t)]\rho(0)$ with ~\cite{alicki06}
\beqar
\label{eq:main:K2}
K_\alpha(t)\rho(0) &=&\sum_{\omega ,\omega ^{\prime }}\int_{0}^{t}e^{i(\omega' -\omega)u}du\int_{-u}^{t-u}F_\alpha(\tau )e^{i\omega \tau }d\tau 
\\ \nonumber
&\times&
\left( S_\alpha(\omega)\rho S_\alpha^{\dag }(\omega')- \frac{1}{2}\{S_\alpha(\omega)S_\alpha^{\dag }(\omega'),\rho \} \right)  
\eeqar
where $F_\alpha (\tau)= \mathrm{Tr}(\rho_{\alpha}R_{\alpha}(\tau)R_{\alpha})$ is the $\alpha$th bath correlation function, $\rho_\alpha$ is the state of the $\alpha$th bath, and $S_{\alpha}(\omega)$ are the Fourier decomposition of the operator $S_{\alpha}$ which are expressed using the Bohr frequencies of $H_0$ through 
\beq
S_\alpha(\omega)=\sum_{n,m}\Pi^{(0)}_n S_{\alpha} \, \Pi^{(0)}_m \delta (\hbar \omega-E^{(0)}_m+E^{(0)}_n),
\eeq 
where $\Pi^{(0)}_n=\ketbra{\psi_n^{(0)}}{\psi_n^{(0)}}$ are the projection operators onto the energy eigenspace of $H_0$. The sums over $\omega$ and $\omega ^{\prime}$ index the system Bohr frequencies. Written in this form, it is clear that the system's eigenstructure is encoded in the propagator. For details see App.~\ref{app:LGKS}.

Equation~(\ref{eq:main:K2}) is a generic result of second order perturbation theory in the system-bath coupling. To bring it to the LGKS form, two additional approximations are carried out. The first approximation assumes that the integral on the left-hand side samples the function $F(\tau)$ in sufficient accuracy to justify the Fourier transform on the right-hand side,
\beqar
  \int_{-u}^{t-u}F_\alpha(\tau )e^{i\omega \tau }d\tau &\approx & \int_{-\infty }^{\infty }F_\alpha(\tau )e^{i\omega \tau }d\tau,
\eeqar
which defines a time independent relaxation rate $\Gamma_\alpha(\omega)$
\beqar
 \Gamma_\alpha(\omega)= \lambda_\alpha^2 \int_{-\infty }^{\infty }F_\alpha(\tau )e^{i\omega \tau }d\tau \geq 0 .
 \label{eq:main:rep1}
 \eeqar
This is the standard Markovian approximation and is valid for long times  such that $\omega t \gg 1$.
The second assumption is typically a stronger condition than the first,
\beqar
\int_{0}^{t}e^{i(\omega'-\omega )u}du &\approx & t\delta _{\omega
\omega'},
\label{eq:main:rep2}
\eeqar
and is referred to as the secular approximation \cite{breuer}. This approximation is  valid when $\min \{ t |\omega-\omega'| \} \gg 1$.
The resultant  Markovian master equation in the Schrodinger picture reads~\footnote{Here and in the rest of the manuscript we neglected the Lamb-type shift correction to the Hamiltonian, that can easily be recovered, see \cite{breuer}.},
\beq
\label{eq:LGKS_ME}
\dot{\rho}=-\frac{i}{\hbar} [H_{0},\rho]+\sum_\alpha \mathcal{D}^\alpha\rho,
\eeq
with the dissipative part
\beqar
\label{eq:dissipator}
\mathcal{D}^\alpha\rho&&=\\ 
&&\sum_{\omega}\Gamma_\alpha(\omega)\left(S_\alpha(\omega)\rho S_\alpha^{\dagger}(\omega)-\frac{1}{2}\left\{ S_\alpha^{\dagger}(\omega)S_\alpha(\omega),\rho\right\} \right) \nonumber
\eeqar
and the relations $S_\alpha(-\omega)=S_\alpha^{\dagger}(\omega)$ and $\Gamma(-\omega)~=~e^{-\beta \hbar\omega}\Gamma(\omega)$. For Bosonic bath at thermal equilibrium, one can factorize the relaxation rate, $\Gamma(\omega)=\gamma(\omega)(n+1)$ and $\Gamma(-\omega)~=~\gamma(\omega)n$, where  $n$ is the Bose-Einstein distribution and $\gamma(\omega)$  is determined by the coupling strength and the density of modes of the bath.   
Note that the simple additive structure for multiple baths in Eq.~(\ref{eq:dissipator}) results from linearity of the original Hamiltonian and the assumption of no initial correlations between the baths.
 
 \section{Response theory within the LGKS framework} 
\label{sec:RespT}
Now we consider adding a constant perturbation at time $t=0$ to the system, which drives the system away from its current NESS to a new steady state. 
 Our treatment of the perturbation $\delta V$ does not constrain it to be the smallest energy scale in the problem. 
It is considered small with respect to the system ``bare'' Hamiltonian $H_0$; however, it is not necessarily small with respect to the system-bath couplings, $\{\lambda_\alpha \}$, which is assumed to be weak. An exact LGKS master equation describing the new dynamics would require deriving the master equation that corresponds to the new Hamiltonian $H_0+\delta V$ from first principles. In many cases, this global approach is convoluted or even impossible. Here, we present an approach based on a perturbative treatment of the LGKS equation in such a way that is consistent with the approximations in Eqs.~(\ref{eq:main:rep1}) and (\ref{eq:main:rep2}), namely, with the secular and Markovian approximations. 
 
The perturbative treatment begins by expanding the eigenvalues and eigenstates of the system Hamiltonian and the perturbation, $H_{0}+\delta V$, in the basis of $H_0$ and in orders of $\delta$,
\beqar
\label{eq:perturbation_expen}
E_{n} & = &E_{n}^{(0)}+\delta E_{n}^{(1)}+\delta^{2}E_{n}^{(2)}+\cdots ,\\ \nonumber
\ket{\psi_{n}} & = &\ket{\psi_{n}^{(0)}}+\delta \ket{\psi_{n}^{(1)}}+\delta^{2}\ket{\psi_{n}^{(2)}}+\cdots.
\eeqar
Given the structure in Eq.~(\ref{eq:main:K2}), the expansion suggests that the perturbation may influence the master equation in two ways: 1) modifying the structure of the master equation due to changes in the \emph{eigenvectors} and 2) modifying the Bohr frequencies of the system and the structure of the master equation, due to changes in the \emph{eigenvalues}.  These two modifications can appear either separately  or together, depending on the timescales of the problem and the details of the perturbation. 

\subsection{Response due to changes in the eigenvectors}
\label{sec:eigenvector}
To start, we consider how changes to the eigenvectors impact the system's response.  
We assume that the Bohr frequencies $\omega$ do not change as a consequence of the perturbation. 
This happens anytime the perturbation shifts all the energy levels by a constant, or more commonly, when the perturbation does not affect the eigenvalues at a given order of $\delta$. 
For example, whenever the eigenvectors of the Hamiltonian $H_0$, and the perturbation $V$ are orthogonal, the first order correction of the eigenvalues vanishes, $E_n^{(1)}=0$.   When only the eigenvectors change, the response theory takes on a transitional perturbative form.

Using Eq.~(\ref{eq:perturbation_expen}), the projection operators can be expressed in orders of $\delta$ where $\Pi^0=\ketbra{\psi_n}{\psi_n}$, and the expansion
\beq
\label{eq:expansion_S}
S(\omega)~=~S^{(0)}(\omega)+\delta S^{(1)}(\omega)+\cdots,
\eeq 
immediately follows.
By virtue of Eq.~(\ref{eq:dissipator}) and (\ref{eq:expansion_S}), the expansion of the dissipator for each bath in orders of $\delta$, 
\beq
\label{eq:expansion_D}
\mathcal{D}\rho=\mathcal{D}_0 \rho+\delta\mathcal{D}_1\rho + \delta^2\mathcal{D}_2 \rho+\cdots,
\eeq 
is analogously clear. 
Here, $\mathcal{D}_j$ is  the  $\delta^j$ order correction to the dynamics,
and the zero order term $\mathcal{D}_0$ corresponds to the dissipator of the unperturbed system with the Hamiltonian $H_0$. 
More specifically, the explicit form of the first order corrections is given by
\beq
\label{eq:S_1}
S^{(1)}(\omega)=\ketbra{\psi_n^{(0)}}{\psi_n^{(1)}}S\ketbra{\psi_n^{'(0)}}{\psi_n^{'(0)}} + \text{similar terms},
\eeq
and
\beqar
\label{eq:D_1}
\mathcal{D}_1\rho=\sum_{\omega}\Gamma(\omega)\left(S^{(0)}(\omega)\rho S^{(1)\dagger}(\omega)\right.
\\ \nonumber
\left. -\frac{1}{2}\left\{ S^{(1)\dagger}(\omega)S^{(0)}(\omega),\rho\right\} \right)+\text{H.c.} .
\eeqar
Similar terms in Eq.~(\ref{eq:S_1}) refers to all terms contributing to first order in $\delta$. 
The procedure above provides a scheme to treat perturbatively complex Hamiltonians within the LGKS formalism. It is simplified significantly by the constant frequencies entering into the sum, which thus do not alter the underlying Markovian or secular approximations.  

The ability to write the expansion Eq.~(\ref{eq:expansion_D}), admits a quantum response theory for NESS to be developed in an analogous way for a closed quantum system. 
After the perturbation is
turned on, the dynamics of a system in contact with multiple baths follows the master equation,
\beq
\label{eq:expansion_ME}
\dot{\rho}=-\frac{i}{\hbar} [H_{0}+\delta V,\rho]+\sum_{\alpha,j=0} \delta^j \mathcal{D}_j^\alpha \rho
\eeq
where for each $\alpha$th bath we can expand the dissipator in a series. 
The new perturbed state of the system $\rho$ can similarly be expanded in orders of $\delta$ as well,
\beq
\label{eq:expansion_rho}
\rho(t)=\pi_{0}(0)+\sum_{j=1} \delta^j \rho_{j}(t)
\eeq
where at time $t=0$ the NESS is $\rho(0)=\pi_0$. 
Note that the $\rho_i$ corrections are not proper density matrices in themselves and satisfy $\tr{}{\rho_j}=0$ at each order.  

Inserting Eq.~(\ref{eq:expansion_rho}) into Eq.~(\ref{eq:expansion_ME}) and keeping only the first order in $\delta$ terms, we have
\beqar
\dot{\rho}_{1}(t) & = &\mathcal{L}_{1}\pi_{0}+\mathcal{L}_{0}\rho_{1}(t),
 \\ \nonumber
\mathcal{L}_{1}\pi_{0} & =& - \frac{i}{\hbar} [V,\pi_{0}]+\sum_\alpha \mathcal{D}^\alpha_1\pi_{0},
\\ \nonumber
\mathcal{L}_{0}\rho_{1}(t) & = & -\frac{i}{\hbar} [H_{0},\rho_{1}(t)]+\sum_\alpha \mathcal{D}^\alpha_0\rho_{1}(t),
\eeqar
with the formal solution  
\beq
\label{eq:rho1}
\rho_{1}(t)  = \int_{0}^{t}d\tau e^{\mathcal{L}_{0}\tau}\mathcal{L}_{1}\pi_{0}.
\eeq
for the first order correction to the density matrix in $\delta$. 
Then, to first order in $\delta$ for an arbitrary observable $A$, 
\beq
\ave A_{\delta}=\ave A+\delta\int d\tau\phi_{A}^{(1)}
\eeq
where the average $\ave{\cdot}_{\delta}$ is taken with respect to the perturbed state. This implies a response function, 
\beqar
\label{eq:response_functions}
\phi_{A}^{(1)} & = &\ave{\mathcal{L}_{1}^{\dagger}A(\tau)} \\
&=&\frac{i}{\hbar} \ave{[V,A(\tau)]}+\sum_\alpha \ave{\mathcal{D}^{\alpha \dagger}_1A(\tau)} \nonumber,
\eeqar
 where $A(\tau)$ is the observable propagated according to the unperturbed dynamics, $A(t)=\Lambda^{\dagger}_0(t)A\equiv e^{\mathcal{L}^{\dagger}_{0}t}A$. Here $\Lambda_0^{\dagger}$ is the dynamical map in the Heisenberg picture with the adjoint LGKS generator $\mathcal{L}_0^{\dagger}$ \cite{breuer}.   
 
Note that this result also holds for an arbitrary initial product steady-state, including thermal equilibrium, in which $\pi_0\propto \exp[-\beta H_0]$, with the inverse temperature $\beta=k_B T$ and the Boltzmann factor $k_B$. Extensions to second and higher order are straight-forward under the assumption that the eigenvalues do not change.

The response functions derived here are different form those found in the literature,  as they involve a term due to the rotation of the dissipator into the perturbed eigenvectors $\mathcal{D}_1$. 
The linear response function now generally has two contributions, $\phi_A^{(1)}=\phi_A^{(1,1)}+\phi_A^{(1,2)}$, with
\beqar
\label{eq:response_split}
\phi_A^{(1,1)}&=&\frac{i}{\hbar} \ave{[V,A(\tau)]}
\\ \nonumber
\phi_A^{(1,2)}&=&\ave{\mathcal{D}\dg_1A(\tau)} \, ,
\eeqar
where the first term $\phi_A^{(1,1)}$ is expected from Kubo theory. The  additional term $\phi_A^{(1,2)}$    originates from  noncommutativity of the new Hamiltonian including the perturbation and the system-bath interaction Hamiltonian, and has no classical counterpart.

For closed quantum systems at thermal equilibrium, which follow a unitary evolution, the term $\phi_{A}^{(1,2)}$ vanishes and the  response function $\phi_{A}^{(1)}$ in Eq.~(\ref{eq:response_functions}) reduces to the standard Kubo formula Eq.~(\ref{eq:Kubo}). 
However, for open quantum systems this is no longer the case. In Sec.~\ref{sec:example1}, we show that this additional contribution can become significant and cannot be neglected.

 We point out that a response function which explicitly includes the dissipative part has been introduced before in the literature~\cite{venuti2016dynamical,albert2016geometry,villegas2016application,pan2020non}. However, in these studies the LGKS-based Kubo
formula is derived under the assumption that a Dyson
expansion of the full evolution exists, without providing a microscopic derivation, or that the perturbation itself is assumed to be represented by a non-Hermitian operator.
By contrast, in this study we show that an Hermitian (Hamiltonian) perturbation of an open quantum system, in the limits discussed above, results in an additional contribution $\phi_A^{(1,2)}$ to the Kubo formula. Moreover, we provide a method to evaluate this term  explicitly  from first Hamiltonian consideration.

\subsection{Example: Two coupled anharmonic oscillators}
\label{sec:example1}
\begin{figure}[t]
\includegraphics[width=8.5cm]{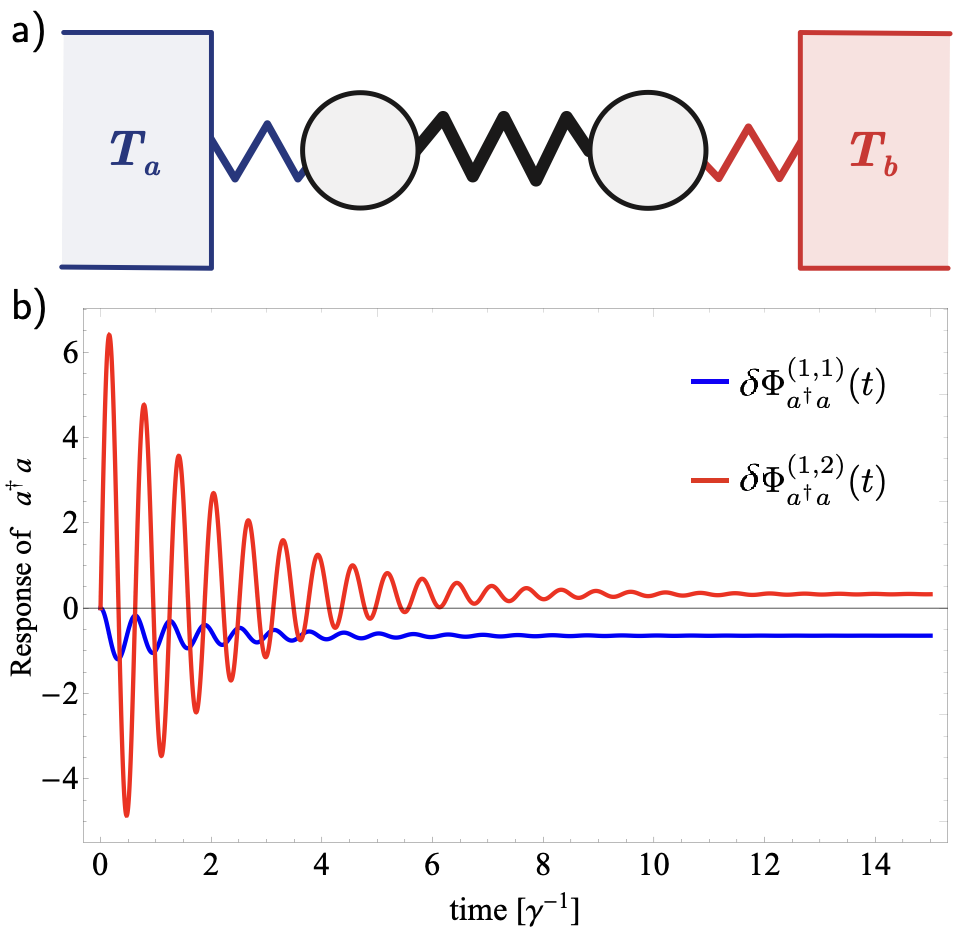}
	\caption{Response of the $a$ oscillator population. a) Schematic illustration of the system of two coupled oscillators. b) The response of the  occupation $a\dg a$ in time. The blue and red lines are the contributions of integrating   Eq.~(\ref{eq:response_F_HO_a}) and Eq.~(\ref{eq:response_F_HO_b}) respectively. The parameters are $\gamma\m=\gamma$,  $\omega=100\gamma$, $g/\hbar=90\gamma$,  $\varepsilon/\hbar=5\gamma$, $\delta=0.02$, $T_a=2\cdot 10^3\hbar \gamma/ k_B$, and $T_b=2.1\cdot 10^3\hbar \gamma/k_B$. For these parameters, according to Eq.~(\ref{eq:ada}), the unperturbed population  $\ave{\ad a}=107.4$.   } 
	\label{fig:response}
\end{figure}

In this section, we illustrate how a perturbation that changes only the eigenvectors of the system to first order results in a linear response determined by both the traditional Kubo term and the additional force from the baths.  
In particular, we  investigate  
the contribution of the different terms  
of $\phi_A^{(1)}$ in Eq.~(\ref{eq:response_split}), illustrating the conditions under which the term $\phi_A^{(1,2)}$ has non-negligible contributions. 
The system we consider consists of two coupled anharmonic oscillators, $a$ and $b$, subject to a linear external field.
Each of the oscillators are coupled to a thermal bath with temperature $T_a$ and $T_b$, respectively as shown schematically in Fig.~\ref{fig:response}a).
The anharmonicity is described by a cubic potential $ \delta V$, which will be considered the perturbation and is treated within the linear response theory.

The unperturbed system Hamiltonian  
\beqar
H_0 &=& \hbar \omega (\ad a + \bd b) +g(\ad b+ a \bd)
\\ \nonumber
&+& \frac{ \varepsilon}{\sqrt{2}}(\ad+a)+\frac{ \varepsilon}{\sqrt{2}}(\bd+b),
\eeqar 
includes the self Hamiltonians of the two harmonic oscillators, a coupling between the modes, and a linear field. Here, $\omega$ denotes the frequency of both oscillators, $g$ their bilinear coupling, and $\varepsilon$ the linear potential that both feel. 
The system-bath interaction Hamiltonians are assumed to be bilinear, 
\beq
H_{sb}~=~ (a + a^{\dagger})\otimes R_a +  (b + b^{\dagger})\otimes R_b \, , 
\eeq
with $R_\alpha = \sum_j \lambda_{\alpha,j} (r_{\alpha,j} + r_{\alpha,j}^{\dagger})$ being a sum of displacement operators of the $\alpha$th bosonic bath, and $\lambda_{\alpha,j}$ the characteristic interaction scale.

The joint master equation for the harmonic system including the linear field is derived using a global approach \cite{levy14b}, which assumes that the inter-coupling is strong compared to the system-bath couplings. The resultant adjoint propagator is
\beqar
\mathcal{L}\dg_0(O) &=& \frac{i}{\hbar}[H_0,O]+\sum_{\ell=\pm, \alpha=a,b}  \mathcal{D}_{0,\ell}^{\alpha \dagger}(O)
+\mathcal{\tilde{D}}_{0,+}^{\alpha \dagger}(O)
\eeqar 
 with the Hamiltonian $H_0=\omega_+ d\p^{\dagger}d\p + \omega_- d\m^{\dagger}d\m$ and the dissipators expressed in a normal mode transformation, $d\p = (a+b)/\sqrt{2}$ and $d\m = (b-a)/\sqrt{2}$, and frequencies $ \omega_{\pm}= \omega\pm g/ \hbar$,
\beqar
\mathcal{D}_{0,\ell}^{\alpha\dagger}(O)&=&\frac{\gamma_\ell(n_\ell^\alpha+1)}{2} \left(d_\ell\dg O d_\ell -\tfrac{1}{2}\{d_\ell \dg d_\ell,O\} \right) \\ \nonumber
&&+\frac{\gamma_\ell n_\ell^\alpha}{2} \left(d_\ell O d_\ell\dg -\tfrac{1}{2}\{d_\ell  d_\ell \dg,O\} \right)
\\ \nonumber
\mathcal{\tilde{D}}_{0,+}^{\alpha \dagger}(O)&=&
\frac{\varepsilon\gamma\p(n\p^\alpha+1)}{2\hbar \omega_+} \left(d\p\dg O+O d\p -\tfrac{1}{2}\{d\p\dg+ d\p,O\} \right)\\ \nonumber
&&+\frac{\varepsilon\gamma\p n\p^\alpha}{2\hbar \omega_+} \left(d\p O +O d\p\dg -\tfrac{1}{2}\{d\p + d\p \dg,O\} \right)  \\ \nonumber
\eeqar  
where we define the 
Bose-Einstein distribution $n_\ell^\alpha~=~[\exp(\beta_\alpha \hbar \omega_\ell)-1]^{-1}$ for each bosonic bath at inverse temperature $\beta_\alpha$, and the decay rates $\gamma_\ell $. The term $\tilde{\mathcal{D}}_{0,+}^{\alpha,\dagger}$ arises from the linear field and can be derived using the perturbation theory that was introduced above (see App.~\ref{app:anharmonic_oscillators} for details).

We will consider for concreteness the response of the occupation number of oscillator $a$ to a step perturbation of the form $ \delta V$, with 
\beq
V =\tfrac{\Omega}{\sqrt{8}}\left(\ad+a-(\bd+b)\right)^3,
\eeq
which adds a cubic potential rendering the system anharmonic expressible as a sum of the new normal modes. Since $\delta$ sets the scale of the perturbation we are free to chose $\Omega \equiv \hbar \omega\m$. For this choice, the perturbative treatment implies $\delta \ll 1$.   In the absence of the perturbation, the occupation number can be calculated using the master equation,
\beq
\label{eq:ada}
\ave{\ad a}=\frac{1}{4}\left(\sum_{l=\pm, \alpha=a,b} n_\ell^\alpha +2\left(\frac{\varepsilon}{\hbar \omega\p}\right)^2 \right)
\eeq
which is valid within an arbitrary steady state.

Following the approach described in previous sections, we find that the first order correction to the eigenvalues vanish, and thus the Bohr frequencies remain the same. The eigenvectors however are modified, leading to corrections of the dissipator in the format of Eq.~(\ref{eq:D_1}) (see App.~\ref{app:anharmonic_oscillators} for details).
To calculate the response function $\phi_{a\dg a}^{(1)}$ of the observable  $\ad a$, we first calculate the evolution of the operator $a\dg a(\tau)=e^{\mathcal{L}\dg_0\tau}a\dg a$  and then evaluate $\ave{\mathcal{L}_1^{\dagger}\ad a(\tau)}$. 
In App.~\ref{app:anharmonic_oscillators}, we show that the time dependent operator $\ad a(\tau)$ can be expressed explicitly as a linear combination of operators at time zero multiplied by time dependent functions.

To study the contribution from the different terms $\propto \delta$, we use the splitting of the response function in Eq.~(\ref{eq:response_split}).
Direct calculation results in
\beq
\label{eq:response_F_HO_a}
\phi_{\ad a}^{(1,1)}= - \frac{3 \varepsilon \omega\m}{\hbar \omega\p} \left (2 \bar{n}\m+1\right ) \sin(\omega\m \tau)e^{-\frac{\gamma\m\tau}{2}},
\eeq
and
\begin{align}
\label{eq:response_F_HO_b}
\phi_{\ad a}^{(1,2)} &&= \frac{3\varepsilon\gamma\m}{2 \hbar \omega\p}\left(\bar{n}\m-1 \right)\left(2 \bar{n}\m+1 \right)\cos(\omega\m \tau)e^{-\frac{\gamma\m\tau}{2}} \, ,
\end{align}
where both contributions depend on the average Bose distribution, $\bar{n}\m=(n^a\m+n^b\m)/2$.  The response functions scale linearly with $\varepsilon$. When no linear field is present, the linear response will vanish completely. Notice that only $\phi_{\ad a}^{(1,2)}$ depends on the kinetic parameters describing the coupling to the bath, similar to the frenetic contributions arising in classically driven diffusive systems~\cite{gao2019nonlinear,lesnicki2020field}.

In Fig.~\ref{fig:response} we plot the  two contributions to the response of the occupation $\ad a$ in time, i.e. the integrated response functions 
\beq
\Phi_{\ad a}^{(1,i)} (t) = \int_{0}^{t}d\tau \, \phi_{\ad a}^{(1,i)}(\tau)
\eeq
for $i=1,2$. In the high temperature limit illustrated in Fig.~\ref{fig:response}, it is clear that the term involving $\phi_{\ad a}^{(1,2)}$ has non-negligible contributions both in the transient and steady-state regimes.  

\begin{figure}[t]
\includegraphics[width=8.5cm]{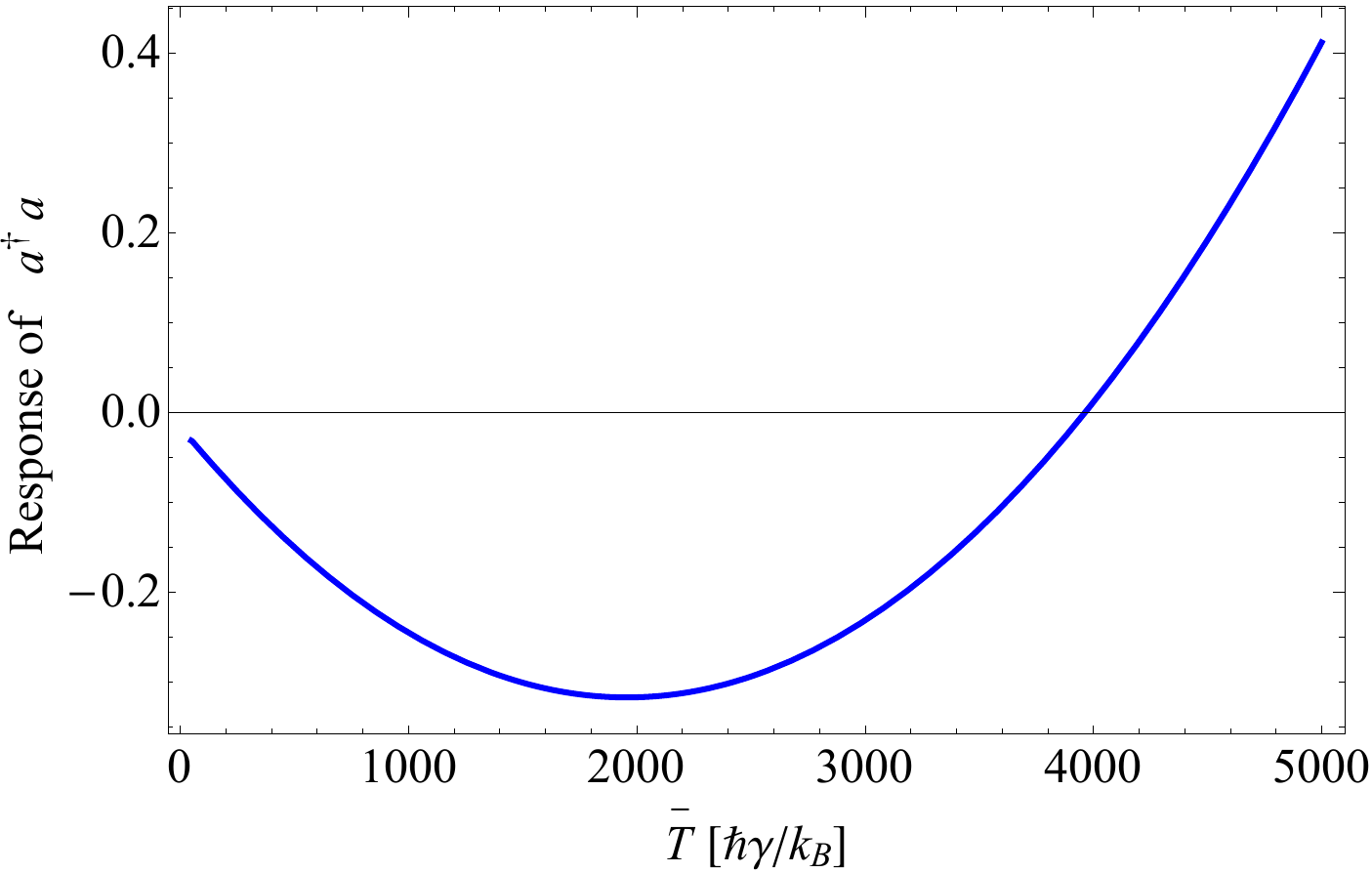}
	\caption{The steady-state response $\delta(\Phi_{\ad a}^{(1,1)}+\Phi_{\ad a}^{(1,2)})$ as a function of the average temperature $\bar{T}$. The parameters are $\gamma\m=\gamma$,  $\omega=100\gamma$, $g/\hbar=90\gamma$, $\varepsilon/\hbar=5\gamma$, $\delta=0.02$ and the temperature difference is $\Delta T=50\hbar \gamma/k_B$.   }  
	\label{fig:response2}
\end{figure}

The significance of this term is also manifested in the scaling behavior of the steady-state response of $\ad a$ with the average temperature $\bar{T}=(T_a+T_b)/2$.
Integrating the response functions over the interval $t=[0,\infty]$,  
\beq
\label{eq:response_HO_a}
\lim_{t\rightarrow \infty} \Phi_{\ad a}^{(1,1)}= - \frac{3 \varepsilon}{\hbar \omega\p} \left( \frac{4\omega\m^2}{\gamma\m^2 +4\omega\m^2}\right)\left(2 \bar{n}\m+1 \right)
\eeq
\begin{align}
\label{eq:response_HO_b}
\lim_{t\rightarrow \infty} \Phi_{\ad a}^{(1,2)}&&=\frac{3\varepsilon}{\hbar \omega\p} \left( \frac{\gamma\m^2}{\gamma\m^2 +4\omega\m^2}\right) \left(\bar{n}\m-1 \right)\left(2 \bar{n}\m+1 \right) \, \nonumber
\end{align}
yields the steady state response. In the high temperature limit, $n_-^\alpha \propto T_\alpha$, which implies that the term $\phi_{\ad a}^{(1,1)} \propto \bar{T}$, whereas $\phi_{\ad a}^{(1,2)} \propto \bar{T}^2$.
This change in the scaling with the average temperature is depicted in Fig.~\ref{fig:response2}, where we plot the total steady state  response, $\delta(\Phi_{\ad a}^{(1,1)}+\Phi_{\ad a}^{(1,2)})$, as a function of the average temperature. As $\bar{T}$ increases, the non-Kubo term, $\Phi_{\ad a}^{(1,2)}$, dominates the behavior with a different scaling and a different sign of the response emergeing.

We note that at high occupation number of the oscillator, the perturbation theory breaks down as the perturbed state depends explicitly on the level number $n$. In this case there is an interplay between the magnitude of $\delta$ and $n$. In particular, we wish $\delta \ll n^{-\frac{1}{2}}$ (see App.~\ref{app:anharmonic_oscillators}).   

Interestingly, at zero temperature, when $n\m^a=n\m^b=0$, the steady-state response
\beq
\Phi_{\ad a}^{(1,1)}+\Phi_{\ad a}^{(1,2)} \rightarrow -\frac{3\varepsilon}{\hbar \omega_+ }
\eeq
is independent of $\gamma_-$, which characterizes the bath and its coupling to the system.  

\subsection{Response due to changes in the eigenvalues}
\label{sec:eigenvalues}
Changes to the eigenvalues $E_n$ leads to modifications of the Bohr frequencies, which can be expanded in terms of $\delta$ as well, $\omega=\omega^{(0)}+\delta \omega^{(1)}+\cdots$.
The procedure introduced for deriving the LGKS master equation can now be repeated using the new perturbed Bohr frequencies and the replacement of the projectors, $\Pi_n^{(0)} \rightarrow \Pi_n$ to a desired order.
 Modifications to the Bohr frequencies introduce new timescales that become relevant for the LGKS approximations in Eqs.~(\ref{eq:main:rep1}) and (\ref{eq:main:rep2}) to be carried out.  
The weak system-bath coupling  limit \cite{davies74,alicki06} that leads to the LGKS master equation assumes a time coarse-graining of fast oscillating terms with frequencies $\nu=\omega'-\omega$. 
This approximation is valid only when the system's relaxation time $\tau$ satisfies $\min \{ \tau |\nu|\}  \gg 1$.
The perturbation may introduce new frequencies, for example at first order in $\delta$,  $\tau |\nu^{(1)}| \gg 1$, terms oscillating with frequencies $\nu^{(1)}= \delta(\omega'^{(1)}-\omega^{(1)})$ are eliminated.  
 Physically, the coarse-granining is carried out when the width of the spectral line is smaller than the level splitting that is now proportional to $\delta$.      

When the Bohr frequencies are modified by the perturbation, expansion of the dissipator in orders of $\delta$ does not necessarily exist. The Fourier transform of the bath correlation functions in Eqs.~(\ref{eq:main:rep1}) and (\ref{eq:dissipator})  includes the perturbed Bohr frequencies $\Gamma_\alpha(\omega^{(0)})\rightarrow \Gamma_\alpha(\delta\omega^{(1)})$. 
This implies that a first-order correction to the eigenvalues, given the coarse-graining discussed above, already results in all orders of the dissipator, making the response of the open system nonlinear. 
Moreover, in the limit $\delta\rightarrow 0$ the dissipator will generally be different from its unperturbed form. The reason is the time coarse-graining, which eliminates terms oscillating with frequencies $\nu^{(1)}$ that are proportional to $\delta$. This discontinuity is a known phenomena when deriving the LGKS master equation from first principle \cite{levy212,alicki89}.  The observations above suggest when the perturbation is small with respect to the system Hamiltonian $H_0$, but  large with respect to the system-bath interactions, response theory for open quantum systems cannot be formulated by standard means. The response function can no longer be associated with correlation functions in the unperturbed system, and requires knowledge of the perturbed state of the system. 

However, when the perturbation is also \textit{small} with respect to the system-bath coupling a generalized fluctuation-dissipation relation can be derived. 
When in addition the first order correction to the eigenvectors vanishes, a local master equation, in which the perturbation does not influence the dissipator, can be applied. In this limit, when the perturbation is small with respect to the system-bath coupling, $\nu^{(1)} \ll \tau^{-1}$, following the development in Sec.~\ref{sec:eigenvector},
\beqar
\phi_{A}^{(1)} & = &i/ \hbar \ave{[V,A(\tau)]},
\eeqar
where contributions from a nonlocal dissipator, that previously were expressed through $\phi_{A}^{(1,2)}$  enters as a higher order cross term, and thus in this instance should be neglected. So although the response function cannot always be associated with correlation function in the unperturbed system, a perturbative treatment based on Eq.~(\ref{eq:perturbation_expen}) offers a practical method  for treating  complex perturbations which cannot be analytically solved exactly. 

\subsection{Example: Two coupled qubits}
\label{sec:example2}
  
 In this section, we study an example illustrating how a perturbation that modifies only the eigenvalues to first order, and does not change the eigenvectors, leads to an LGKS master equation that involves higher orders of the perturbation as discussed in the previous section. 
While the perturbation is assumed to be small compared to the system Hamiltonian, the response of the system is investigated in two different limits.
The first assumes the perturbation is small compared to the system-bath couplings, which in terms of timescale can be translated to $ \nu^{(1)} \ll \tau^{-1}$, with $\tau$ the typical relaxation time of the system. 
Since the system-bath coupling is already assumed to be weak, the regime of applicability is quite restricted and a local approach can be implemented.
The opposing limit, in which the perturbation is assumed to be large with respect to the system-baths coupling, i.e. $ \nu^{(1)} \gg \tau^{-1}$, leads to new Bohr frequencies and to the coarse-grained dynamics discussed above.     

\begin{figure}[t]
	\includegraphics[width=8.5cm]{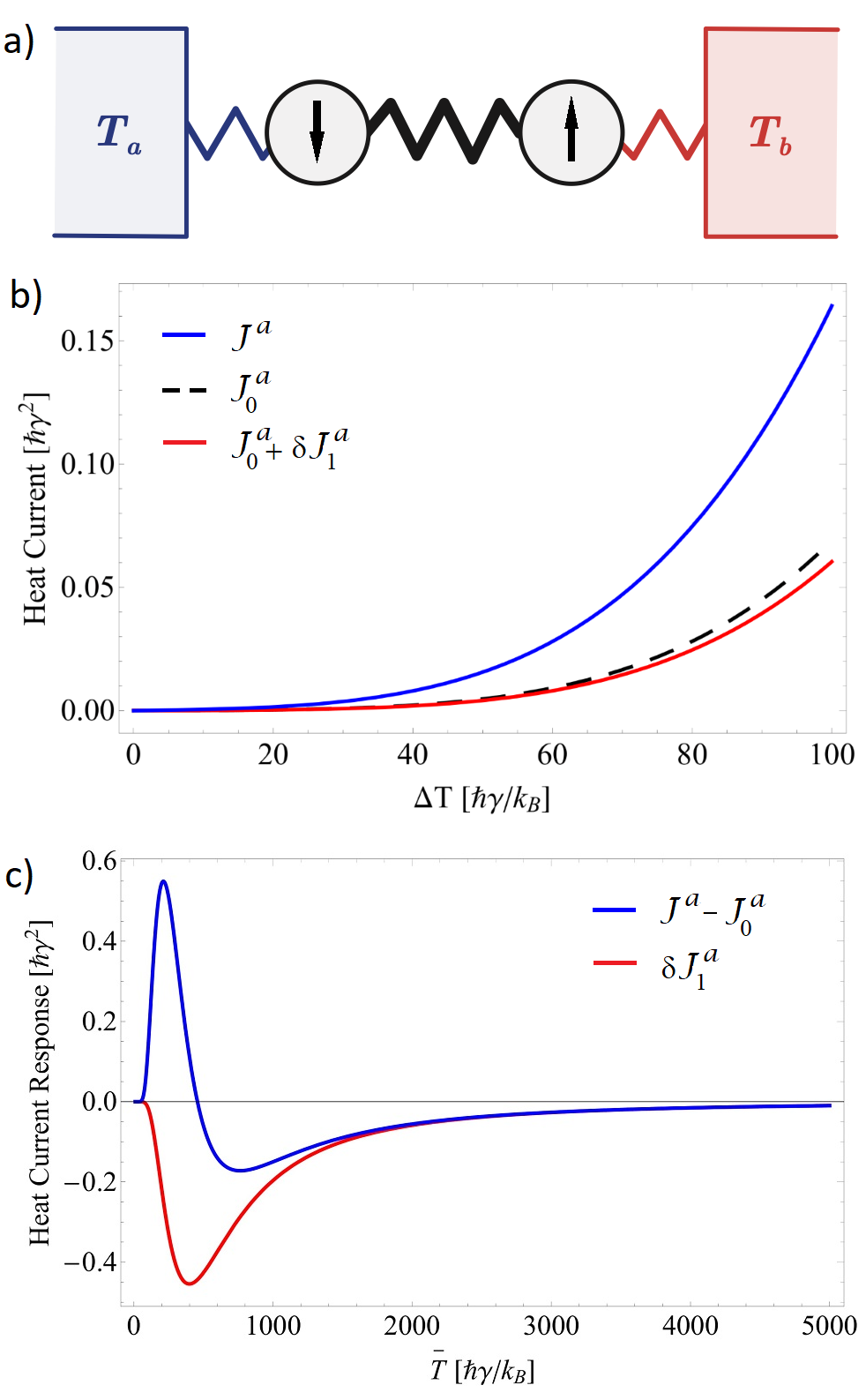}
	\caption{  a) Schematic illustration of the system. b) The steady-state heat currents with  and without (black) the perturbation at low temperatures, $\bar{T}=50\hbar\gamma/k_B$, as function of the temperature difference between the baths. The red line corresponds to $\mathcal{J}_0^a + \delta \mathcal{J}_1^a$, and the blue line to $\mathcal{J}^a$ of Eq.~(\ref{eq:J_global_TLS}).  c)  The steady-state heat current response as a function of the average temperature $\bar{T}$ for a fixed temperature difference $\Delta T=20\hbar \gamma/k_B$ . In red, the first-order correction $\delta\mathcal{J}_1^a$. In blue, the difference, $\mathcal{J}^a-\mathcal{J}^a_0$, between the new and old steady-state heat current (subtracting Eq.~(\ref{eq:J0_TLS}) from Eq.~(\ref{eq:J_global_TLS}). The parameters are     $\gamma\m=\gamma\p=\gamma$, $\omega=10^3\gamma$, $g/\hbar=2\cdot 10^2 \gamma$, , and $\delta \Omega /\hbar=50\gamma$.}
\label{fig:qubitsresponse}
\end{figure}

In this example, the response of the heat current between the thermal baths and a finite size system is studied. In particular,
we consider a chain of two qubits $A$ and $B$ that are coupled to two bosonic heat-baths with temperature $T_a$ and $T_b$, respectively. A schematic of the system is shown in Fig.~\ref{fig:qubitsresponse}a).
The unperturbed system Hamiltonian, including the qubits and the coupling between them, reads    
\beq
\label{eq:H0_qubits}
H_0=\frac{\hbar \omega}{2} \left ( \sigma_z^A + \sigma_z^B \right )+\frac{g}{2}(\sigma_x^A\sigma_x^B+\sigma_y^A\sigma_y^B) \, ,
\eeq
where $\sigma_i$ is a Pauli matrix, $\omega$ is the transition frequency of the uncoupled qubits, and $g$ the coupling between them. The system-bath coupling Hamiltonian is assumed to be bilinear,  
\beqar
H_{s b} = \sigma_x^A\otimes R_a + \sigma_x^B\otimes R_b
\eeqar
where $R_\chi=\sum_k \lambda_{\chi,k}(r_{\chi,k} +r^{\dagger}_{\alpha,k})$ with $\chi=\{a,b\}$ is the weighted displacement operator for the $\chi$th bath.
In App.~\ref{app:qubits} we derive the global  master equation which is used to calculate the NESS properties of the system, noting that the eigenvalues of $H_0$ are $\omega_{\pm}=\omega\pm g/\hbar$ each with a twofold degeneracy. The resultant adjoint propagator is given by,
\beq
\label{eq:master_equation_qubits}
\mathcal{L}\dg_0 (O) = \frac{i}{\hbar}[H_0,O]+ \sum_{\ell=\pm, \chi=a,b} \mathcal{D}_{0,\ell}^{\chi \dagger}(O),
\eeq
with
\beqar
\mathcal{D}_{0,\ell}^{\chi \dagger}(O)&=& \gamma_\ell(n_\ell^\chi+1)\left(\chi\dg_\ell O \chi_\ell-\frac{1}{2}\{\chi_\ell \chi_\ell \dg,O\} \right)
\\ \nonumber
&+&
\gamma_\ell n_\ell^\chi\left(\chi_\ell O \chi\dg_\ell-\frac{1}{2}\{\chi\dg_\ell \chi_\ell,O\} \right).
\eeqar
where again $n_{l}^\chi=(\exp(\beta_\chi \hbar \omega_\ell)-1)^{-1}$ is the Bose-Einstein distribution at inverse temperature $\beta_\chi$ for the $\chi$th bath.
The operators appearing in $\mathcal{D}_{0,\ell}^{\chi}$ satisfy the relation $[H_0,\chi_{\pm}]=-\hbar \omega_{\pm}\chi_{\pm}$ with $\chi_\pm$ being either $
a_\pm ~=~ (\sigma\m^A \mp \sigma_z^A\sigma\m^B)/2$ or $b_\pm ~=~(\sigma\m^B \mp \sigma\m^A\sigma_z^B)/2$. Each channel has a decay rate $\gamma_\ell $.

We consider the response of the heat flow through the system at steady-state, $\mathcal{J}$, to a perturbation of the form 
\beq
 V= \delta \Omega \sigma_z^A \sigma_z^B,
\eeq
with $\Omega$ the coupling energy. At first order, this perturbation changes the eigenvalues, leaving the eigenvectors the same. The average heat flow from the $a$ bath in the absence of the perturbation is given by $
\mathcal{J}_0^a ~=~\sum_{\ell=\pm}\ave{\mathcal{D}_{0,\ell}^{ a \dagger } H_0}$ (see Eq.~(\ref{eq:heat_flow}) and App.~\ref{app:qubits}). At steady state,
\beq
\label{eq:J0_TLS}
\mathcal{J}_0^a=\frac{\gamma\p \hbar \omega\p(n_{+}^a-n_{+}^b)}{4\left(n_{+}^a+n_{+}^b+1 \right)} +\frac{\gamma\m \hbar \omega\m(n_{-}^a-n_{-}^b)}{4\left(n_{-}^a+n_{-}^b+1) \right)}
\eeq
we find heat flows in two channels with energies $\hbar\omega_{\pm}$ and corresponding rates proportional to $\gamma_{\pm}$. In the high temperature limit, the heat current  $\mathcal{J}^a_0 \propto \Delta T$ scales as the temperature difference  $\Delta T~=~T_a-T_b$. 

When the perturbation is much smaller than the coupling to the baths, $ \nu^{(1)} \ll \tau^{-1}$, a local master equation with system Hamiltonian $H_0+\delta V$ in the commutator and  a local dissipative part $\mathcal{D}_0$ can be justified.
 The basis of this approximation is neglecting terms of the order $O( \delta^2 \lambda^2)$ and higher in the master equation, and 
 the fact that the first-order correction of the eigenvectors vanishes.
Incorporating such terms in the master equation would require introducing higher order corrections in the system-bath coupling for consistency.
 
The heat current from the $a$-bath, to first order, is given by $\mathcal{J}_0^a +  \delta \mathcal{J}_1^a$, where $\mathcal{J}_0^a$ is given by Eq.(\ref{eq:J0_TLS}), and  
$
\mathcal{J}_1^a ~=~\sum_{\ell=\pm}\ave{\mathcal{D}_{0,\ell}^{\dagger a } V}$, which is strictly the local contribution of the perturbation. Here we wish to emphasize that locality does not refer to individual qubits, as these are treated globally (see Eq.~(\ref{eq:J0_TLS})), but rather to the perturbation itself. 
 At steady-state, 
\beq
\label{eq:J1_TLS}
\mathcal{J}_1^a = -\frac{\gamma\m \Omega (n\m^a-n\m^b)+\gamma\p \Omega (n\p^a-n\p^b)}{2(n\m^a+ n\m^b+1)(n\p^a+ n\p^b+1)}.
\eeq
Note that $\mathcal{J}_1^a$ has the opposite sign has that of $\mathcal{J}_0^a$, and the overall contribution is determined by the sign of $\delta$. 
At the high temperature limit, $\mathcal{J}_0^a~\propto~\Delta T/\bar{T}$, whereas $\mathcal{J}_1^a~\propto~\Delta T/\bar{T}^2$ with $\bar{T}=(T_a+T_b)/2$ the average temperature. 

In the opposite limit where the perturbation is much larger than the coupling to the bath, $ \nu^{(1)} \gg \tau^{-1}$, the baths act to thermalize  a different system with a different spectrum. To first order in $\delta$ the eigenvalues and the corresponding eigenvectors read
\begin{align*}
&\{ E_{00}=-\hbar \omega+\delta\Omega, E\pm =\pm g-\delta\Omega, E_{11}=\hbar\omega+\delta\Omega\}
\\ \nonumber
&\{\ket{00}, \ket{\pm}=\dfrac{1}{\sqrt{2}}(\ket{01}\pm\ket{10}), \ket{11}\} 
\end{align*}
which in this case also happens to be the exact correction for any $\delta$. 

 The Fourier decomposition of the interaction Hamiltonian operators can be expressed as
\beqar
\sigma_x^A &=& S_{1}+S_{2}+S_{3}+S_{4} + H.c.
\\ \nonumber
\sigma_x^B &=& -S_{1}+S_{2}-S_{3}+S_{4} + H.c.,
\eeqar  
with the corresponding frequencies 
\beqar
\label{eq:S_qubit}
S_{1}&=& \frac{1}{\sqrt{2}}\ketbra{00}{-};\quad  \omega_{1}=\omega\m-2\delta\Omega/\hbar
\\ \nonumber
S_{2}&=&\frac{1}{\sqrt{2}}\ketbra{00}{+};\quad \omega_{2}=\omega\p-2\delta\Omega/\hbar
\\ \nonumber
S_{3}&=& \frac{1}{\sqrt{2}}\ketbra{+}{11};
\quad \omega_{3}=\omega\m+2\delta\Omega/\hbar
\\ \nonumber
S_{4}&=&\frac{1}{\sqrt{2}}\ketbra{-}{11};
\quad \omega_{4}=\omega\p+2\delta\Omega/\hbar.
\eeqar

\begin{figure*}
\centering
\subfloat{\includegraphics[width=0.45\linewidth]{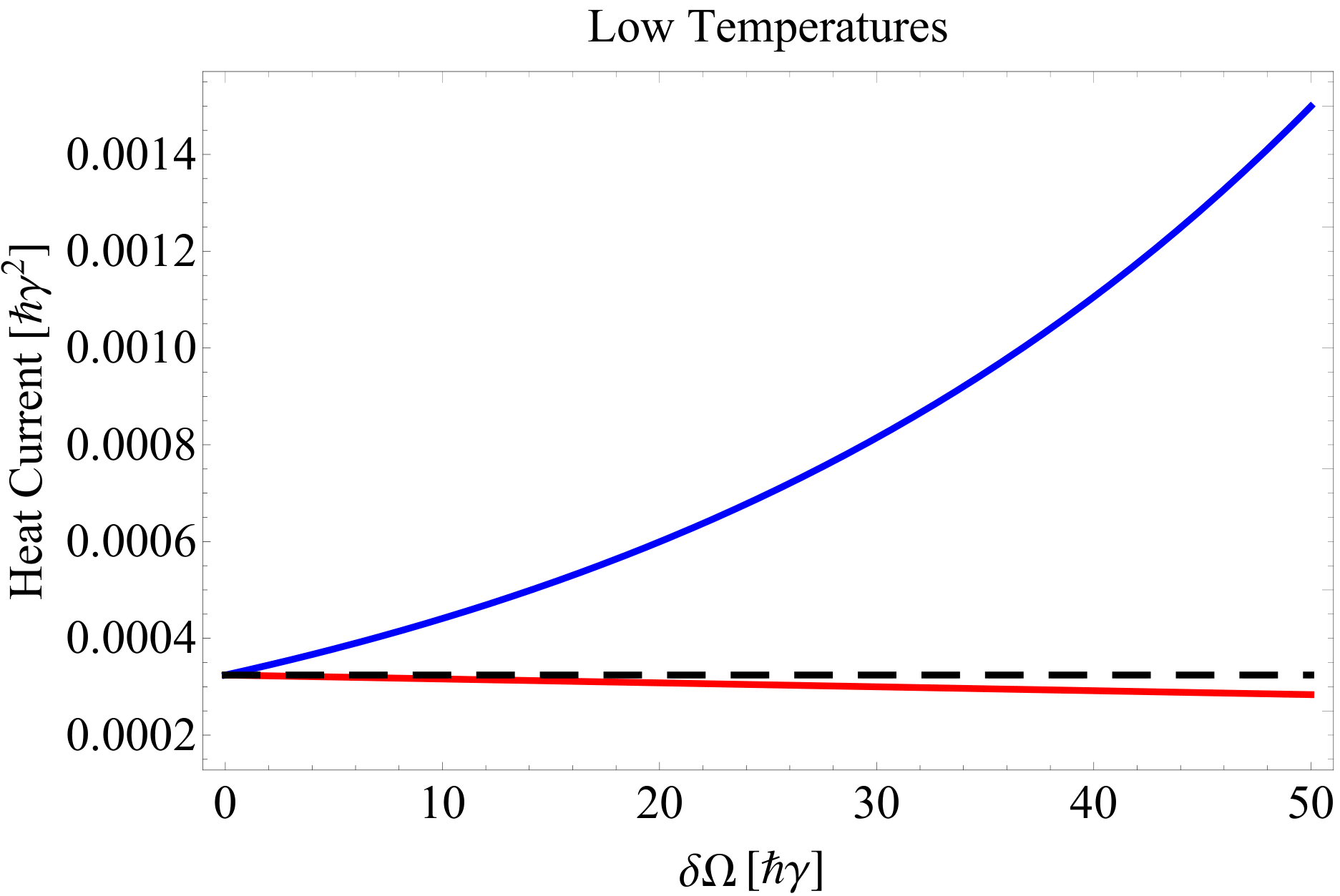}}
\qquad
\subfloat{\includegraphics[width=0.45\linewidth]{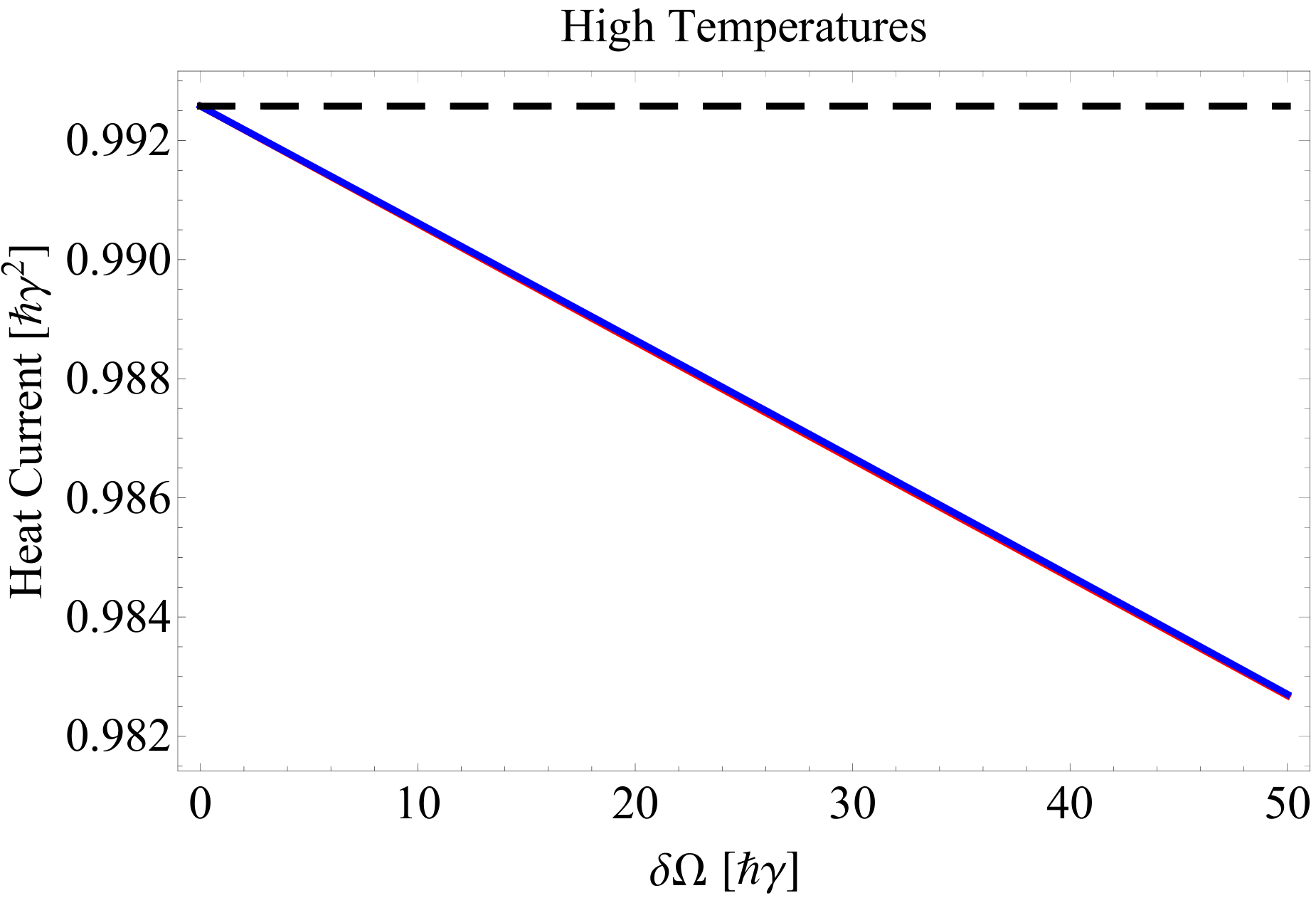}}
	\caption{ The steady-state heat current as function of the perturbation strength. In black, the heat current of the unperturbed system given by Eq.~(\ref{eq:J0_TLS}). In red, the sum  $\mathcal{J}_0^a+ \delta J_1^a$ of Eqs.~(\ref{eq:J0_TLS}) and (\ref{eq:J1_TLS}) which correspond to the limit $ \nu^{(1)}\ll \gamma$. In blue, the heat current given by Eq.~(\ref{eq:J_global_TLS}) in the apposing limit $ \nu^{(1)}\gg\gamma$. The left caption correspond to a low temperature limit with the average temperature $\bar{T}=50\hbar\gamma/k_B$, where as the right caption correspond to high temperature limit with $\bar{T}=5\cdot 10^3\hbar\gamma/k_B$. The parameters $\gamma\m=\gamma\p=\gamma$, $\omega=10^3 \gamma$, $g=2\cdot10^2\gamma$ and $\Delta T= 20\hbar\gamma/k_B$.   } 
	\label{fig:qubitsresponse2}
\end{figure*}

Following this decomposition a master equation can be derived (see App.~\ref{app:qubits}), and the heat current from the $a$-bath reads
\beqar
\label{eq:J_global_TLS}
\mathcal{J}^a&=&\sum_{j=1}^4 \ave{\mathcal{D}_j^a H}_{\rho}
\\ \nonumber
 &=&\sum_{j=1}^4 \hbar \omega_j\left( -\gamma_j(n^a_j+1)\ave{S\dg_j S_j}_{\rho} + \gamma_j n^a_j\ave{S_j S\dg_j}_{\rho} \right) \, .
\eeqar
 An analytic expression of Eq.~(\ref{eq:J_global_TLS}) exists, however, it is to involved to present here. 
Note that averages in Eq.~(\ref{eq:J_global_TLS}) are taken with respect to the new nonequilibrium state. 
While for the unperturbed system heat was flowing in two channels, in the current limit, heat flows in four channels that correspond to the perturbed Bohr frequencies $\{\omega_j\}$ of Eq.~(\ref{eq:S_qubit}).

Figure~\ref{fig:qubitsresponse}b) illustrates the steady-state heat currents with and without the perturbation at low temperatures and as function of the temperature difference between the two baths. The black dashed line corresponds to the heat flow of the unperturbed system $\mathcal{J}_0^a$. The red and blue lines indicate the heat flow subject to perturbation calculated when $ \nu^{(1)} \ll \tau^{-1}$ and $ \nu^{(1)} \gg \tau^{-1}$, respectively. Note that according to Eq.~(\ref{eq:J_global_TLS}) we have $\nu^{(1)}=4 \delta\Omega/\hbar$. When first order corrections to the Bohr frequencies are accounted for, we observe a  significant amplification of the steady-state heat current (blue line). Whereas, standard response theory predicts only very small changes to the heat flow. 

An informative comparison between the limits is obtained by plotting        
the response of the heat current at steady-state as a function of the average temperature $\bar{T}$ Fig.~\ref{fig:qubitsresponse}c).
The red line corresponds to the first-order correction $ \delta \mathcal{J}_1^a$, and the blue to the difference between the new and old steady-state heat flow, $\mathcal{J}^a- \mathcal{J}_0^a$, of Eqs.~(\ref{eq:J0_TLS}) and (\ref{eq:J_global_TLS}). 
At low temperatures, the behavior of the heat current response in the two limits is substantially different. 
This  clearly indicates that the perturbation has a significant effect on the dissipation caused by the coupling to the baths.  
While in both cases the perturbation is considered small with respect to the system Hamiltonian, its relation to the relaxation rate results in a different response behavior. 
 In the high temperature limit, the correction to the Bohr frequencies becomes negligible, the heat current responses in the two limits coincide and vanish asymptotically, and the relaxation rate increases with the temperature.

The fact that a first order perturbation, with respect to the system Hamiltonian, can lead to a nonlinear response of the heat current is illustrated in Fig.~(\ref{fig:qubitsresponse2}). In this figure we plot the steady-state heat current as function of the perturbation strength $\delta$. The black dashed line represent the heat current of the unperturbed system given by Eq.~(\ref{eq:J0_TLS}), the red line is the sum  $\mathcal{J}_0^a+ \delta J_1^a$ of Eqs.~(\ref{eq:J0_TLS}) and (\ref{eq:J1_TLS}) that correspond to the limit $\nu^{(1)}\ll \tau^{-1}$, and the blue line is the heat current given by Eq.~(\ref{eq:J_global_TLS}) in the apposing limit $\nu^{(1)}\gg \tau^{-1}$. 
In the low temperature limit, small changes in the eigenvalues lead to pronounced and nonlinear changes in the heat current with respect to the unperturbed heat current.
In the high temperature limit, as may be expected, the perturbation has very little effect on the heat current, as it is dominated by the  temperature.         

\section{Heat flows, entropy and  entropy production}
\label{sec:heatf}
As stressed by Alicki and others \cite{alicki89,alicki06,levy14b,trushechkin2016perturbative}, although LGKS operators cannot generically  be associated with an underling Hamiltonian, a thermodynamic consistency requires this association, as a consistent microscopic description of dissipation is not otherwise possible.
To accomplish this faithfully, the LGKS operators requires information on the eigenstrcture of the system whose transitions they promote, as encoded by the detail balance relationship between their transition rates. As we have shown above, to accurately obey these constraints in the presence of a perturbation requires care.

To illustrate the thermodynamic compliance of the theory discussed above, we introduce the response of the heat flows, entropy and entropy production.
While Eq.~(\ref{eq:response_functions}) 
provides the response function of a general quantum observable of the system, in this section we introduce the response of thermodynamic functions which are state dependent. 
The entropy production $\sigma$ has contributions from changes in the internal entropy of the system and from the exchange of heat with the baths at different temperatures
\beq
\label{eq:entropy_prod}
\sigma(t)=\frac{d}{dt}S(t) -\sum_j \frac{\mathcal{J}^\alpha(t)}{T_\alpha}.
\eeq 
Here $S(t)=-k_B\tr{}{\rho(t)\ln\rho(t)}$ is the von-Neumann entropy and $\mathcal{J}^\alpha$ is the heat flow from the $\alpha$-bath
\beq
\label{eq:heat_flow}
\mathcal{J}^\alpha(t)=-\beta_j^{-1} \ave{\mathcal{D}^{\alpha \dagger} \ln \pi^\alpha}_{\rho}=\ave{\mathcal{D}^{\alpha \dagger} H}_\rho.
\eeq 
The state  $\pi^{\alpha}$ appearing in Eq.~(\ref{eq:heat_flow})  is the thermal state of the generator $\mathcal{D}^\alpha $, with the  temperature $\beta_\alpha^{-1}=k_B T_\alpha$. 
The average, on the other hand, is taken with respect to the time dependent state, $\ave{\cdot}_{\rho}=\tr{}{\rho(t) \cdot}$.
Noting that
\beq
 \frac{d}{dt}S(t)=-k_B\tr{}{(\mathcal{L} \rho(t))\ln \rho(t)}, 
\eeq 
using $\mathcal{D}=\sum_\alpha \mathcal{D}^\alpha$, Eqs.~(\ref{eq:entropy_prod}) and (\ref{eq:heat_flow}) together with Spohn's inequality 
$
-\tr{}{(\mathcal{L}\rho)(\ln\rho-\ln\pi )} \geq 0
$ \cite{spohn78}, 
we conclude that the entropy production is non-negative $\sigma(t)\geq 0$ (see App.~\ref{app:entropy_production_and_heat_flow}). 
 In case the eigenvalues are perturbed, as discussed in Sec.~\ref{sec:eigenvalues}, the modified LGKS generator of each bath has a unique Gibbs-like  stationary state that corresponds to the perturbed spectrum.
This is demonstrated explicitly in the example of Sec.~\ref{sec:example2} in which the perturbation treatment and the global approach coincide.  
Following similar arguments discussed above, the perturbed entropy production is nonnegative as well. This is detailed in Apps.~\ref{subsec:app_C2} and \ref{subsec:app_D1}. 
 We mention in passing that if particles are also exchanged between the baths and the system, one needs to account for the heat carried by them, see for example \cite{thingna2017kinetics}. 

Next, we derive the linear response for the case discussed in Sec.~\ref{sec:eigenvector}, when the perturbation changes only the eigenvectors. 
At time $t=0$, before the perturbation is applied, the steady state heat flow from the $\alpha$-bath is strictly given by  $\mathcal{J}_0^\alpha\,=\,-\beta_\alpha^{-1}\ave{\mathcal{D}_0^{\alpha \dagger} \ln \pi_0^\alpha}$. 
The first order correction, $\mathcal{J}_1^\alpha~=~\partial_\delta \mathcal{J}^\alpha|_{\delta=0}$, can be expressed in two useful forms. The first reads
\beqar
\label{eq:J1_a}
\mathcal{J}_1^\alpha (t)&=&
-\beta_\alpha^{-1} \int_0^t \ave{\mathcal{L}_1^{ \dagger} \Lambda_0^{\dagger}(\tau)\mathcal{D}_0^{\alpha \dagger} \ln \pi_0^\alpha} d\tau 
\\ \nonumber
&-&\beta_\alpha^{-1}\left( \ave{\mathcal{D}_1^{\alpha\dagger}\ln \pi_0^\alpha}+\ave{\mathcal{D}^{\alpha \dagger}_0 \partial_{\delta}\ln\pi^\alpha|_{\delta=0}}\right),
\eeqar
where $\pi^\alpha\propto e^{-\beta_\alpha(H_0+\delta V)}$ is a unique stationary state of the  generator $\mathcal{L}^\alpha$,
see App.~\ref{app:entropy_production_and_heat_flow} for details. The second form, 
\beqar
\label{eq:J1_b}
\\ \nonumber
\mathcal{J}^\alpha_1(t)&=&
\int_0^t \ave{\mathcal{L}_1^{ \dagger} \Lambda_0^{\dagger}(\tau)\mathcal{D}_0^{\alpha \dagger} H_0} d\tau + \ave{\mathcal{D}_1^{\alpha\dagger}H_0}+\ave{\mathcal{D}^{\alpha \dagger}_0 V},
\eeqar
shows clearly that the sum of the heat flows from all the baths returns the total energy change in the system. 

The first order correction of the von-Neumann entropy, 
\beq
\label{eq:S1}
S_1=\partial_{\delta}S|_{\delta=0}=-k_B\tr{}{\rho_1\ln \pi_0},
\eeq
allows us to calculate the correction for the change in the internal entropy, see App.~\ref{app:entropy_production_and_heat_flow},
\beqar
\label{eq:dS1dt}
\nonumber
\frac{d}{dt}S_1(t)&=&-k_B\left(\tr{}{(\mathcal{L}_0\rho_1)\ln\pi_0}+\tr{}{\pi_0 \mathcal{L}\dg_1\ln\pi_0} \right)
\\ \nonumber
&=& -k_B\int_{0}^t \ave{\mathcal{L}_1^{\dagger}\Lambda_0^{\dagger}(\tau)\mathcal{L}_0^{\dagger}\ln\pi_0}d\tau -k_B\ave{\mathcal{L}_1^{\dagger}\ln\pi_0 }
\\ 
&=& -k_B \ave{\mathcal{L}_1^{\dagger} \Lambda_0^{\dagger}(t)\ln \pi_0}.
\eeqar
In the second equality, we applied Eq.~(\ref{eq:rho1}), and in the last, we integrated using the relations $\Lambda_0^{\dagger}(\tau)\mathcal{L}_0^{\dagger}~=~\frac{\partial}{\partial \tau}\Lambda_0^{\dagger}(\tau)$, and $\Lambda_0^{\dagger}(0)=1$.
Assuming the existence of a new stationary state, one may observe that, as may be expected, for $t\rightarrow\infty$ the derivative $\frac{d}{dt}S_1=0$. To show this, we assume $\rho_1(\infty)=\pi_1$ such that $\mathcal{L}(\pi_0+\pi_1)=O(\delta^2)$. Then, $\mathcal{L}_0\pi_1=-\mathcal{L}_1\pi_0$, which implies that for sufficiently long  times, the first line in Eq.~(\ref{eq:dS1dt}) vanishes.

At equilibrium,  the first order correction to the entropy production vanishes at all times $\sigma_1(t)=0$. The change in the von-Neumann entropy equals the entropy flow from the bath. For systems at NESS, this is no longer the case. The correction can turn negative or positive.

\section{outlook}     
\label{sec:outlook}
Obtaining an accurate description of  open quantum systems' response to perturbation is imperative for understanding the dynamics  and manipulating the outcome and performance of quantum devices. 
In this study, we employed a
microscopic Hamiltonian  approach based on physical arguments, rather than a phenomenological approach, to develop a quantum response theory. 
This distinction is crucial, as one cannot associate an arbitrary Hamiltonian with a given dissipator of the dynamics. 
The perturbation modifies the system and the way it is perceived by the surroundings. As a result, the dissipator is modified, and in the linear response, corrections to the standard response function $\phi_A^{(1,1)}$ are already expected.

Combining stationary Hamiltonian perturbation theory with the LGKS formalism, we developed a useful practical scheme to obtain the response functions for both systems at equilibrium or NESS. 
The approach reveals the role of changes in the eigenvaectors and eigenvelues as a result of the perturbation. 
Eigenvector modifications lead to corrections of the dissipator that can be expressed order by order in the perturbation parameter. It  therefore allows us to derive response functions that include new non-negligible terms. 
For closed quantum systems at thermal equilibrium, our result recovers the standard Kubo formula.
The new terms can be interpreted as a response to forces induced by the surroundings as a result of the perturbation. 

Changes in the eigenvalues  lead to modifications of the Bohr frequencies, and in turn to modifications of the operators appearing in the dissipator.
In this scenario, a distinction between two limits is appropriate. First, when the system-bath coupling is small compared to the perturbation, linear changes to the eigenvalues lead to the system's nonlinear response. Although now the response function cannot be expressed by correlation functions at equilibrium or NESS, the perturbative approach introduces a genuine scheme treating complex Hamiltonians within the LGKS formalism. 

In the opposing limit, the coarse-graining procedure that leads to the LGKS master equation is not justified. 
However, when the eigenvectors are not influenced by the perturbation and since the perturbation is a assumed to be small compared to the system-bath coupling, a local master equation is justified and the linear response is strictly given by $\phi_A^{(1,1)}$ of Eq.~(\ref{eq:response_split}). 
Taking into account higher-order contributions would make sense only if higher orders in the system-baths coupling are considered as well. While in this work we focused on stationary perturbations, there is still a need for a quantum response theory of time-dependent perturbations that are derivable from a physical Hamiltonian perspective. 

\section*{Acknowledgements}
We thank Wenjie Dou for stimulating discussions. E.R. acknowledges support from the Department of Energy, Photonics at Thermodynamic Limits Energy Frontier Research Center, under Grant DE-SC0019140. 
DTL was supported by the U.S. Department of Energy, Office of Science, Basic Energy Sciences, CPIMS Program Early Career Research Program under Award No. DE-FOA-0002019.

\appendix
\section{ Thermalizing LGKS master equation in the weak-coupling limit} 
\label{app:LGKS}
We briefly review the main ingredients for deriving the mater equation in the weak coupling limit. 
More elaborate derivations can be found in Refs.\cite{davies74,alicki87,breuer}.
We assume a quantum system with the Hamiltonian $H_0=\sum_m E_m\ketbra{m}{m}$ coupled to a thermal bath $\rho_R$ with the Hamiltonian $H_R$ such that $[\rho_R,H_R]=0$, via an interaction Hamiltonian $ H_I=\lambda S\otimes R$. Here, $\lambda$ represents the coupling strength which is assumed small, and $S$ and $R$ are linear operators of the system and bath respectively. The generalization to a system coupled to several bath is straight  forward and follow the linearization of the master equation for initially uncorrelated baths.

Working in the interaction picture, the reduced dynamics of the system is  given by the partial trace over the bath degrees of freedom  
 \begin{equation}
\rho (t)=\Lambda (t)\rho \equiv \mathrm{Tr}_R \bigl(U_{I}(t)\rho\otimes\rho_R U_{I}(t)^{\dagger}\bigr),
\label{eq:ap:red_dyn}
\end{equation}
where
\begin{equation}
U_{I}(t) = \mathcal{T}\exp\Bigl\{-i\lambda\int_0^t S(s)\otimes R(s)\,ds\Bigr\},
\label{eq:ap:eprop_int}
\end{equation}
is the time ordered propagator in the interaction picture, which is defined using 
\begin{equation}
S(t) = e^{iH_0t/\hbar} S e^{-iH_0t/\hbar},\quad \  R(t)= e^{iH_R t/\hbar} R e^{-i/H_R t/\hbar}.
\label{eq:ap:prop_int1}
\end{equation}

As shown in \cite{alicki06,levy212}, the dynamical map $\Lambda(t)$ can be expressed by the cumulant expansion
\begin{equation}
\Lambda (t)=\exp \sum_{n=1}^{\infty }[\lambda ^{n}K^{(n)}(t)].
\end{equation}
Since we assume the state of the bath is thermal, then $\tr{}{\rho_R R}=0$,   which implies that the first order cumulant $K^{(1)}=0$.
The Born approximation (weak coupling) consists of terminating
the cumulant expansion at $n=2$, henceforth we denote $K^{(2)}\equiv K$ and
\beq
\Lambda (t)=\exp [\lambda ^{2}K(t)+O(\lambda ^{3})].
\eeq
One  obtains%
\beqar
K(t)\rho &=&\int_{0}^{t}ds\int_{0}^{t}du F(s-u)\times
\\ \nonumber
 & &\left( S(s)\rho S^{\dag }(u)- \frac{1}{2}\{S(s)S^{\dag }(u),\rho \} \right),
\label{eq:K(t)}
\eeqar
with the bath correlations $F(s)= \mathrm{Tr}(\rho _{R}R(s)R)$.  
The Markov approximation (in the interaction picture) means that for long enough time (or short correlation time) one can use the following approximation
\beq
K(t)\simeq t\mathcal{L}  \label{eq:L}
\eeq
where $\mathcal{L}$ is a LGKS generator. To find its form we first decompose
$S(t)$ into its Fourier components, which in the interaction picture reads
\beq
S(t)=\sum_{\omega } e^{-i\omega t}S(\omega) , \quad \quad S(-\omega )= S^{\dagger}(\omega ),
\eeq
where the set $\{\omega=E_n-E_m\}$ contains the \emph{Bohr frequencies} of the system  Hamiltonian.
This decomposition is equivalent to the requirement $[H_0,S(\omega)]=\hbar \omega S(\omega)$.
Expression (\ref{eq:K(t)}) then reads, 
\beqar
\nonumber
K(t)\rho &=&\sum_{\omega ,\omega ^{\prime }}\left( S(\omega)\rho S^{\dag }(\omega')- \frac{1}{2}\{S(\omega)S^{\dag }(\omega'),\rho \} \right) \times
\\ 
& &\int_{0}^{t}e^{i(\omega' -\omega)u}du\int_{-u}^{t-u}F(\tau )e^{i\omega \tau }d\tau.  
\label{eq:K2}
\eeqar
To bring (\ref{eq:K2}) to the LGKS form, two approximations are carried out
\beqar
\nonumber
  \int_{-u}^{t-u}F(\tau )e^{i\omega \tau }d\tau &\approx & \Gamma(\omega)=\int_{-\infty }^{\infty }F(\tau )e^{i\omega \tau }d\tau \geq 0,
\\ 
\int_{0}^{t}e^{i(\omega'-\omega )u}du &\approx & t\delta _{\omega
\omega'}.
\label{eq:rep1}
\eeqar
The first approximation assumes that the integral on the left-hand side sample  the function $F(\tau)$ in sufficient accuracy in order to justify the Fourier transform on the right-hand side. This approximation is valid for long times  such that $t\ll 1/\omega$.
The second assumption is typically a stronger condition than the first, and is referred to as the secular approximation. This approximation is  valid when $t\ll \max \{1/|\omega-\omega'|\}$.

Finally,  the  Markovian master equation in the Schrodinger picture reads 
\begin{eqnarray}
\frac{d\rho }{dt} &=&-\frac{i}{\hbar}[H,\rho ]+\mathcal{D}\rho ,   \\ \notag
\mathcal{D}\rho  &\equiv &\sum_{\omega }\Gamma(\omega) \left( S(\omega)\rho S^{\dag }(\omega)- \frac{1}{2}\{S(\omega)S^{\dag }(\omega),\rho \} \right)  
\label{Dav}
\end{eqnarray}



\par
%
%
%
%
%
%
%
%

\section{Two coupled anharmonic oscillators in a linear field }
\label{app:anharmonic_oscillators}
\subsection{Two coupled harmonic oscillators in a linear field }
We derive the master equation  for the unperturbed system, two coupled harmonic oscillators $A$ and $B$ in a linear field
\beqar
H_0 &=& \hbar \omega\ad a +\hbar \omega \bd b + g(\ad b+ a \bd)
\\ \nonumber
&+& \frac{ \varepsilon}{\sqrt{2}}(\ad+a)+\frac{ \varepsilon}{\sqrt{2}}(\bd+b).
\eeqar
Each of the oscillators is coupled to a thermal Bosonic bath with temperature $T_a$ and $T_b$ respectively. The baths Hamiltonian and its interaction to the system is denoted by
\beqar
H_B&=&\sum_j \hbar \omega_j r\dg_{a,j} r_{a,j}+\sum_j \hbar \omega_j r\dg_{b,j} r_{b,j}
\\ \nonumber
H_{I}&=& (a\dg+a)\sum_j\lambda_{a,j}(r\dg_{a,j}+ r_{a,j})
\\ \nonumber
&+&(b\dg+b)\sum_j\lambda_{b,j}(r\dg_{b,j}+ r_{b,j}).
\eeqar

Performing the transformation 
\beqar
a\rightarrow\tfrac{1}{\sqrt{2}}(d\p -d\m)\\ \nonumber
b\rightarrow\tfrac{1}{\sqrt{2}}(d\p + d\m), 
\eeqar
the Hamiltonian reads
\beq
H_0 =\hbar \omega\p d\p\dg d\p+\hbar \omega\m d\m\dg d\m +  \varepsilon(d\p +d\dg\p),
\eeq
where $\omega_{\pm}=\omega\pm g/\hbar$, the commutators $[d_{\pm},d\dg_{\pm}]=1$, and all other commutation relation are zero.
The system-bath interaction Hamiltonian then reads
\beqar
H_{I}&=& \frac{1}{\sqrt{2}}(d\dg\p +d\p -(d\dg\m +d\m) )\sum_j\lambda_{a,j}(r\dg_{a,j}+ r_{a,j})
\\ \nonumber
&+& \frac{1}{\sqrt{2}}(d\dg\p +d\p +(d\dg\m +d\m) )\sum_j\lambda_{b,j}(r\dg_{b,j}+ r_{b,j}).
\eeqar
Moving to the interaction picture we have
\beqar	
e^{iH_0t/\hbar}d\m e^{-iH_0t/\hbar}&=&d\m e^{-i\omega\m t}
\\ \nonumber
e^{iH_0t/\hbar}d\p e^{-iH_0t/\hbar}&=&\left(d\p-\frac{\varepsilon}{\hbar \omega\p}\right) e^{-i\omega\p t}.
\eeqar 
Next, the derivation of the master equation in the weak coupling limit can now be performed as presented in appendix~\ref{app:LGKS}. 
When the difference between the new Bohr frequencies $|\omega\p -\omega\m|= 2g/\hbar$ is greater than the relaxation rate, terms rotating at that frequency are neglected and the two bosonic modes $d\p$ and $d\m$ can be considered as independent harmonic oscillators, where each is coupled to both baths. 
The joint master equation for the Harmonic system including the linear field is then given by,
\beqar
\mathcal{L}\dg_0(O) &=& i/\hbar[H_0,O]+\sum_{\ell=\pm, \alpha=a,b}  \mathcal{D}_{0,\ell}^{\alpha\dagger}(O)
+\mathcal{\tilde{D}}_{0,+}^{\alpha\dagger}(O)
\\ \nonumber 
\mathcal{D}_{0,\ell}^{\alpha\dagger}(O)&=&\frac{\gamma_\ell(n_\ell^\alpha+1)}{2} \left(d_\ell\dg O d_\ell -\tfrac{1}{2}\{d_\ell \dg d_\ell,O\} \right)
\\ \nonumber
&+&
\frac{\gamma_\ell n_\ell^\alpha}{2} \left(d_\ell O d_\ell\dg -\tfrac{1}{2}\{d_\ell  d_\ell \dg,O\} \right)
\\ \nonumber
\mathcal{\tilde{D}}_{0,+}^{\alpha\dagger}(O)&=&
\frac{\varepsilon\gamma\p(n\p^\alpha+1)}{2\hbar\omega_+} \left(d\p\dg O+O d\p -\tfrac{1}{2}\{d\p\dg+ d\p,O\} \right)
\\ \nonumber
&+&
\frac{\varepsilon\gamma\p n\p^\alpha}{2\hbar\omega_+} \left(d\p O +O d\p\dg -\tfrac{1}{2}\{d\p + d\p \dg,O\} \right) . 
\eeqar  
Here we defined the Bose-Einstein distribution $n_\ell^\alpha=[\exp(\hbar \omega_\ell\beta_\alpha)-1]^{-1}$, and the decay rates $\gamma_\ell\equiv \gamma(\omega_\ell)$,  with $\ell=\pm$ and $\alpha=a,b$.
The term $\tilde{\mathcal{D}}_{0,+}^{\alpha\dagger}$ arising from the linear field, can also be derived using the perturbation theory that was introduced above. 
In this case, a first order perturbation of the linear field recovers the exact master equation, just like a perturbation theory for the harmonic oscillator in a linear field of a closed system.

\subsection{Cubic perturbation}
We study the linear response 
of the system above to a cubic potential of the form
\beq
 \delta V=\tfrac{ \delta\Omega}{\sqrt{8}}\left(\ad+a-(\bd+b)\right)^3.
\eeq 
After the transformation 
\beq
\delta V= \delta\Omega\left(d\dg\m+d\m\right)^3.
\eeq
Setting $\Omega\equiv\hbar\omega\m$, the first order correction for the eigenvalues vanishes, whereas, the eigenvectors
\beqar
\ket{\psi_n} &=& \ket{n}-3\delta\sqrt{(n+1)^3}\ket{n+1}+3\delta\sqrt{n^3}\ket{n-1}
\nonumber \\ 
&-&\frac{\delta}{3}\sqrt{(n+3)(n+2)(n+1)}\ket{n+3}\\ \nonumber
 &+& \frac{\delta}{3}\sqrt{n(n-1)(n-2)}\ket{n-3}+O(\delta^2),
\eeqar
where $\ket{n}$ is the eigenstate of $d^{\dagger}_{-}d_{-}$ .
Since the eigenstates depends explicitly  on the level number $n$, the validity of the perturbation theory at high occupation number  
is determined by the interplay between $n$ and delta. At large $n$, i.e. $n\gg 1$, the lowest order energy correction is proportional  to $\delta^2n^2$ which should be smaller compared to the zero order that scales linearly with $n$, which implies $\delta \ll n^{-\frac{1}{2}}$.

Next, up to first order in $\delta$
\begin{widetext}
\beqar
\ketbra{\psi_n}{\psi_n} =\ketbra{n}{n}&+&\delta \left( 
- 3\sqrt{\frac{(n+1)^3}{8}}\ketbra{n}{n+1}+3\sqrt{n^3}\ketbra{n}{n-1} -\frac{1}{3}\sqrt{(n+3)(n+2)(n+1)}\ketbra{n}{n+3} \right.
 \nonumber \\
&+&\left. 
  \frac{1}{3}\sqrt{n(n-1)(n-2)}\ketbra{n}{n-3}+ \text{h.c.}
	\right)+O(\delta^2).
\eeqar  
\end{widetext}
The Fourier decomposition of the interaction Hamiltonian with frequency $\omega\m$ reads
\beqar
S(\omega\m)&=&\sum_{n} \ketbra{\psi_n}{\psi_n} (d\dg\m + d\m)\ketbra{\psi_{n+1}}{\psi_{n+1}} 
\nonumber \\ \nonumber
 &=& d\m +\delta\left( d\m^{\dagger 2} -3d\m^2 + 6d\dg\m d\m+3 \right)
\\  
 &\equiv & d\m+ \delta J\m, 
\eeqar
and satisfies
\beqar
[H_0+ \delta V,S(\omega\m)]=-\hbar \omega\m S(\omega\m)+O(\delta^2)
\\ \nonumber
[H_0+ \delta V,S\dg(\omega\m)]=\hbar \omega\m S\dg(\omega\m)+O(\delta^2).
\eeqar
Other frequencies will not contribute to the first order.
Keeping order of $\delta$ we arrive at
\beqar
\mathcal{L}_1^{\dagger}(O)&=&i/\hbar[ V,O]+\sum_{\alpha=a,b} \mathcal{D}_{1,-}^{\alpha\dagger}(O)
\\ \nonumber
\mathcal{D}_{1,-}^{\alpha\dagger}(O)&=&\frac{ \gamma_-(n_-^\alpha+1)}{2} \left(d_-\dg O J_- -\tfrac{1}{2}\{d_- \dg J_-,O\} \right)
\\ \nonumber
&+&
\frac{ \gamma_- n_-^\alpha}{2} \left(J_- O d_-\dg -\tfrac{1}{2}\{J_-  d_- \dg,O\} \right)
\\ \nonumber
&+&\frac{\gamma_-(n_-^\alpha+1)}{2} \left(J_-\dg O d_- -\tfrac{1}{2}\{J_- \dg d_-,O\} \right)
\\ \nonumber
&+&
\frac{ \gamma_- n_-^\alpha}{2} \left(d_- O J_-\dg -\tfrac{1}{2}\{d_-  J_- \dg,O\} \right).
\eeqar
Here we defined $J_- = \left( d_-^{\dagger 2}-3d_-^2+6 d_-\dg d_- +3 \right)$ and the relaxation rate $\gamma_-\equiv \gamma(\omega_-)$.
\par

\subsubsection{Response of the observable $a\dg a$ }
To evaluate the response of the observable $\ad a$ to the anharmonicity, we first need to calculate $\ad a(t)$, that takes the analytic form
\beqar
\ad a(t) &=& e^{\mathcal{L}_0\dg t}\ad a \rightarrow  e^{\mathcal{L}_0\dg t} \left(d\p\dg d\p +d\m\dg d\m - u \right)
 \nonumber \\ \nonumber
&=&f_{t}^1 d\m\dg d\m+f_{t}^2 x_- +f_t^3 y_-+f_t^4 u+f_t^5 v 
\\ 
&+& f_t^6 d\p\dg d\p +f_t^7 x_++f_t^8 y_+ + f_t^9. 
\eeqar 
Here $f_t^i$ are time dependent functions and we used the short notation $x_{\pm}=d_{\pm}\dg +d_{\pm}$, $y_{\pm}=i( d_{\pm}-d_{\pm}\dg)$, $u=d\p\dg d\m+ d\p d\m\dg$ and $v=i(d\p\dg d\m- d\p d\m\dg)$. Since  $\mathcal{L}_1\dg$ acts only on the manifold $(-)$, the terms of $f_t^6$ to $f_t^9$ will vanish. In order to calculate 
\beq
\phi_{\ad a}^{(1)}(\tau)=\ave{\mathcal{L}_1\dg \left( \ad a(\tau) \right)},
\eeq
we are left with evaluating
\beqar
\mathcal{L}_1\dg (d\m\dg d\m)&=&-i \left(d\m\dg J\m - J\m\dg d\m + d\m\dg K\m - K\m\dg d\m \right) 
\\ \nonumber
&+&\sum_{\alpha=a,b}\frac{\gamma_-}{4}\left(\left(d\m\dg K\m +K\m\dg d\m \right)  \right.
\\ \nonumber
&-& \left.
 \left(d\m\dg J\m+ J\m\dg d\m \right)-12n_-^\alpha x_- \right)
\\ \nonumber
\mathcal{L}_1\dg (x_-)&=& \sum_{\alpha=a,b}\frac{\gamma_-(n_-^\alpha+1)}{4}(J\m +J\m\dg)-3\gamma_-(2n_-^\alpha+1)
\\ \nonumber
\mathcal{L}_1\dg (y_-)&=&6\omega\m x_-^2 + \sum_{\alpha=a,b}\frac{ \gamma\m(n\m^\alpha+1)}{4}i(J\m - J\m\dg)
\\ \nonumber
\mathcal{L}_1\dg (u)&=&3\omega\m y\p x\m^2 -3\frac{\gamma_-(2n_-^\alpha+1)}{2 }x_+ 
\\ \nonumber
&+& \sum_{\alpha=a,b} \frac{\gamma\m (n_-^\alpha+1)}{4}\left(d\p\dg J\m+d\p J\m\dg \right)
\\ \nonumber
\mathcal{L}_1\dg (v)&=& 3\omega\m x\p x\m^2 + 3\frac{\gamma_-(2n_-^\alpha+1)}{2 }y_+
\\ \nonumber
&+&\sum_{\alpha=a,b} \frac{\gamma\m (n_-^\alpha+1)}{4}i\left(d\p\dg J\m - d\p J\m\dg \right),
\eeqar
with $K\m\equiv 2d\m^{\dagger 2} + 6d\m^2 $. Averaging with respect to the NESS we are left with
\beqar
\ave{\mathcal{L}_1\dg (d\m\dg d\m)}&=&0
\\ \nonumber
\ave{\mathcal{L}_1\dg (x\m)}&=& \sum_{\alpha=a,b}3 \gamma_-(n_-^\alpha+1)\ave{d\m\dg d\m}
\\ \nonumber
& -&\frac{3\gamma_-(3n_-^\alpha+1)}{2}
\\ \nonumber
\ave{\mathcal{L}_1\dg (y\m)}&=& 12\omega\m \ave{d\m\dg d\m}+6\omega\m
\\ \nonumber
\ave{\mathcal{L}_1\dg (u)}&=& \sum_{\alpha=a,b} \frac{6\gamma_-(n_-^\alpha+1)}{4}\ave{x\p}\ave{d\m\dg d\m} 
\\ \nonumber
&-&\frac{3\gamma_-(3n_-^\alpha+1)}{4}\ave{x\p}
\\ \nonumber
\ave{\mathcal{L}_1\dg (v)}&=& 3\omega\m\ave{x\p}\left(2\ave{d\m\dg d\m }+1 \right),
\\ \nonumber
\eeqar
and using the steady state averages
\beqar
\ave{d\m\dg d\m }&=&\frac{ n\m^a+ n\m^b}{2}
\\ \nonumber
\ave{d\p\dg d\p }&=&\frac{ n\m^a+ n\m^b}{2}+\left(\frac{\varepsilon}{\hbar \omega\p}\right)^2
\\ \nonumber
\ave{d\p }&=&-\frac{\varepsilon}{\hbar \omega\p},
\eeqar
we arrive at the result

\beqar
\phi_{\ad a}^{(1)}&=&\phi_{\ad a}^{(1,1)}+\phi_{\ad a}^{(1,2)}
\\ \nonumber
\phi_{\ad a}^{(1,1)}(\tau)&=&f_{\tau}^3 \ave{\mathcal{L}_1\dg (y\m)}
+f_{\tau}^5 \ave{\mathcal{L}_1\dg (v)}
\\ \nonumber
\phi_{\ad a}^{(1,2)}(\tau)&=&f_{\tau}^2 \ave{\mathcal{L}_1\dg (x\m)}
+f_{\tau}^4 \ave{\mathcal{L}_1\dg (u)}.
\eeqar

Where
\beqar
f_{\tau}^2 &=& \frac{\varepsilon}{2\hbar\omega\p} \left(\cos(\omega\m\tau)e^{-\frac{\gamma\m}{2} \tau} -\cos(2g\tau/\hbar)e^{-\frac{\gamma\m+\gamma\p}{2} \tau} \right)
 \nonumber \\ \nonumber
f_{\tau}^3 &=& -\frac{\varepsilon}{2\hbar\omega\p} \left(\sin(\omega\m\tau)e^{-\frac{\gamma\m}{2} \tau} +\sin(2g\tau/\hbar)e^{-\frac{\gamma\m+\gamma\p}{2} \tau} \right)
\\ \nonumber
f_{\tau}^4 &=& -\frac{1}{2} \cos(2g\tau/\hbar)e^{-\frac{\gamma\m+\gamma\p}{2} \tau} 
\\
f_{\tau}^5 &=& -\frac{1}{2} \sin(2g\tau/\hbar)e^{-\frac{\gamma\m+\gamma\p}{2} \tau}. 
\eeqar

\section{Perturbed two coupled qubits}
\label{app:qubits}

\subsection{Unperturbed system}

We derive the master equation for two coupled qubits  with the Hamiltonian Eq.~(\ref{eq:H0_qubits}). Each of the quibits $A$ and $B$ is coupled to a heat-bath with temperature $T_a$ and $T_b$ respectively, and is given by the Hamiltonian  $\sigma_x^A\otimes R_a+\sigma_x^B\otimes R_b$ with $R_\chi=\sum_k \lambda_{\alpha,k}(r_{\chi,k} +r^{\dagger}_{\chi,k})$.
The global master equation reads 

\beq
\label{eq:master_equation_qubits}
\frac{d}{dt}\rho= -i/\hbar[H_0,\rho]+\sum_{\chi=a,b}\sum_{\ell=\pm}\mathcal{D}_{0,\ell}^{\chi}\rho,
\eeq
with
\beqar
\mathcal{D}_{0,\ell}^{\chi}\rho&=& \gamma_j(n_{\ell}^\chi+1)\left(\chi_\ell\rho \chi\dg_\ell-\frac{1}{2}\{\chi_\ell\dg \chi_\ell,\rho\} \right)
\\ \nonumber
&+&
\gamma_\ell n_\ell^\chi\left(\chi_\ell\dg \rho \chi_\ell-\frac{1}{2}\{\chi_\ell \chi_\ell\dg,\rho\} \right).
\eeqar
Here $n_\ell^\chi=(\exp(\beta_\chi \hbar \omega_\ell)-1)^{-1}$, with inverse temperature $\beta_\chi$ of the $\chi$-bath and $\omega_{\pm}=\omega\pm g/\hbar$.
The operators appearing in $\mathcal{D}_{0,l}^{\chi}$ satisfy the relation $[H_0,\chi_{\pm}]=-\hbar \omega_{\pm}\chi_{\pm}$ with $\chi=a,b$ and 
\beqar
a\p &=&(\sigma\m^A-\sigma_z^A\sigma\m^B)/2
\qquad
a\m=(\sigma\m^A+\sigma_z^A\sigma\m^B)/2
\\ \nonumber
b\p&=&(\sigma\m^B-\sigma\m^A\sigma_z^B)/2
\qquad
b\m=(\sigma\m^B+\sigma\m^A\sigma_z^B)/2.
\eeqar

\subsection{The limit $\nu^{(1)} \ll \tau^{-1}$}
\label{subsec:app_C2}
In this limit, oscillating terms with frequencies $4 \delta \Omega/ \hbar$ vanish and the master equation takes the form
\beq
\frac{d}{dt}\rho = -i/\hbar[H_0+\delta V,\rho]+\sum_{i=a,b}\sum_{j=1}^4 \mathcal{D}_j^i \rho,
\eeq
with
\beqar
\mathcal{D}_j^i \rho &=& \gamma_j(n_j^i+1)\left(S_j\rho S\dg_j-\frac{1}{2}\{S\dg_j S_j,\rho\}\right)
\\ \nonumber
&+&\gamma_j n_j^i \left( S\dg_j\rho S_j-\frac{1}{2}\{S_j S\dg_j,\rho\} \right).
\eeqar

 Note that each generator $\mathcal{D}_j^i$ has a unique Gibbs-like stationary state with temperature $T_i$ and frequency $\omega_j$ given in Eq.~(\ref{eq:S_qubit}). According to the proof presented in App.~\ref{subsec:app_D1} it follows that the entropy production of the perturbed system in this example is nonnegative.

\section{Entropy production and heat flows}
\label{app:entropy_production_and_heat_flow}
\subsection{Entropy production and heat flows of a quantum system coupled to multiple baths}
\label{subsec:app_D1}
In the weak coupling limit, energy and entropy flows from the baths can be associated to changes in the energy of the quantum system determined by the partial generators $\mathcal{L}^\alpha$ of the $\alpha$-bath.   
The change in the energy of a quantum system with Hamiltonian $H$ can be written as
\beqar
\ave{\frac{d}{dt}H}&=&\tr{}{\rho\mathcal{L}^{\dagger}H}=\sum_\alpha \tr{}{\rho\mathcal{L}^{\alpha \dagger}H}
\\ \nonumber
&=&\sum_\alpha \tr{}{\rho\mathcal{D}^{\alpha \dagger}H}\equiv \sum_\alpha\mathcal{J}^\alpha
\eeqar   
Since we assume that $\mathcal{L}^\alpha\equiv -i/\hbar[H,\cdot]+\mathcal{D}^\alpha$ is the thermal generator of the $\alpha$-bath, each $\mathcal{L}^\alpha$ as a unique Gibbs like stationary state $\pi^\alpha \propto e^{-\beta_\alpha H}$ and we can write
\beq
\label{eq:J}
\mathcal{J}^\alpha=-k_B T_\alpha \tr{}{\rho\mathcal{L}^{\alpha \dagger} \ln \pi^\alpha}=-k_B T_\alpha \tr{}{(\mathcal{L}^{\alpha }\rho) \ln \pi^\alpha}.
\eeq  
At steady state the sum of all heat flows vanish as expected.
The change in the von-Neumann entropy reads
\beq
\frac{d}{dt} S=-k_B\tr{}{(\mathcal{L}\rho)\ln\rho}
\eeq
Putting these together we see that the entropy production is always positive
\beqar
\label{eq:app_entropy_production}
\sigma(t)&=&\frac{d}{dt} S -\sum_\alpha \frac{\mathcal{J}^\alpha}{T_\alpha}
\\ \nonumber
&=&\sum_\alpha -k_b\tr{}{(\mathcal{L}^\alpha\rho)\ln\rho}+ k_B\tr{}{\mathcal{L}^\alpha\ln \pi^\alpha}\geq 0
\eeqar

The inequality follows from Spohn inequality $-\tr{}{(\mathcal{L}\rho)(\ln\rho-\ln\pi)}\geq 0$ that apply to any LGKS generator with a stationary state $\pi$. Since any $\mathcal{L}^{\alpha}$ has a unique stationary state $\Pi^{\alpha}$ then also the sum over $\alpha$ in Eq.~(\ref{eq:app_entropy_production}) is nonnegative.   

\subsection{First order correction of the heat flows }
The first order corrections for the heat flow can be derived in two ways.  In the first, the starting point is by considering the total energy change of the system. The first order correction  reads
\beqar
\label{eq:ap:energy_change}
\partial_{\delta}\tr{}{\rho\mathcal{L}\dg H}|_{\delta=0}&=&\tr{}{\rho_1\mathcal{L}\dg_0 H_0}+\tr{}{\pi_0\mathcal{L}\dg_1 H_0}
\\ \nonumber
&+&\tr{}{\pi_0\mathcal{L}\dg_0 V}
\\ \nonumber
&=& \ave{\mathcal{L}\dg_1\Lambda\dg_0H_0} + \ave{\mathcal{L}\dg_0 V}
\eeqar 
In the last equality we used eq.(\ref{eq:rho1}) for $\rho_1$, the relation $\partial_\tau \Lambda\dg_0H_0=\Lambda\dg_0\mathcal{L}\dg_0H_0$, and  taking the integral explicitly.
One can also note that at steady state, $\rho_1\rightarrow \pi_1$, the energy change vanishes as expected. This is an immediate consequence of the relation $\mathcal{L}_0\pi_1=-\mathcal{L}_1\pi_0$ which holds at steady state.
  Based on the first equality in (\ref{eq:ap:energy_change}) we can identify the correction from the $\alpha$-bath
\beqar
\label{eq:ap:J1_1}
\mathcal{J}^\alpha_1&=&\tr{}{\rho_1\mathcal{L}^{\alpha\dagger}_0 H_0}+\tr{}{\pi_0\mathcal{L}^{\alpha\dagger}_1 H_0}+\tr{}{\pi_0\mathcal{L}^{\alpha \dagger}_0 V}
\\ \nonumber
&=&-k_B T_\alpha \left(\int_0^t \ave{\mathcal{L}_1^{ \dagger} \Lambda_0^{\dagger}(\tau)\mathcal{L}_0^{\alpha \dagger} \ln \pi_0^\alpha} d\tau + \ave{\mathcal{L}_1^{\alpha\dagger}\ln \pi_0^\alpha}\right)
\\ \nonumber
&+&\ave{\mathcal{L}^{\alpha \dagger}_0 V}
\eeqar

The second approach is by expanding Eq.~(\ref{eq:J}), that gives
\beqar
\label{eq:ap:J1_2}
\mathcal{J}_1^\alpha &=& \partial_\delta \mathcal{J}^\alpha|_{\delta=0}
\\ \nonumber
 &=& -k_B T_\alpha \left(\int_0^t \ave{\mathcal{L}_1^{ \dagger} \Lambda_0^{\dagger}(\tau)\mathcal{L}_0^{\alpha \dagger} \ln \pi_0^\alpha} d\tau + \ave{\mathcal{L}_1^{\alpha\dagger}\ln \pi_0^\alpha}\right)
\\ \nonumber
&-&k_B T_\alpha \ave{\mathcal{L}_0^{\alpha \dagger}\partial_{\delta}\ln\pi^\alpha|_{\delta=0}},
\eeqar
where $\pi^\alpha$ is the new steady state of the generator $\mathcal{L}^\alpha$. Assuming the $\alpha$-bath generator $\mathcal{L}^\alpha$ has a unique stationary state $\pi^\alpha\propto e^{-\beta_\alpha(H_0+\delta V)}$ , equations (\ref{eq:ap:J1_1}) and (\ref{eq:ap:J1_2}) become identical.   

\subsection{First order correction of the entropy and entropy production} 
To prove Eq.(\ref{eq:S1}) for the first order correction  of the von-Neumann entropy, we note that
\beqar	
S_1&=&\partial_{\delta}S|_{\delta=0}=-k_b\tr{}{\rho_1\ln \pi_0}-k_b\tr{}{\rho\frac{d}{d\delta}\ln \rho|_{\delta=0}}
\nonumber \\ 
&=&-k_B\tr{}{\rho_1\ln \pi_0}.
\eeqar
To show that  the term $\tr{}{\rho\frac{d}{d\delta}\ln \rho|_{\delta=0}}=0$, we use the expansion
\beq
\rho(\delta)\frac{d}{d\delta}\ln\rho(\delta)=\frac{d\rho}{d\delta}+\frac{1}{2}\rho^{-1} \left[ \rho, \frac{d\rho}{d\delta}\right]+\frac{1}{3}\rho^{-2} \left[\rho, \left[ \rho, \frac{d\rho}{d\delta}\right]\right]+\cdots ,
\eeq
the trace property, and the fact that $\tr{}{\rho_1}=0$.

Taking the time derivative and keeping first order corrections  we arrive at
\beqar
\label{eq:dS1dt_app}
\frac{d}{dt}S_1(t)&=&-k_B\left(\tr{}{\mathcal{L}_0(\rho_1(t))\ln\pi_0}+\tr{}{\mathcal{L}_1(\pi_0)\ln\pi_0} \right)
 \nonumber \\ \nonumber
&=& -k_B\left(\int_{0}^t \ave{\mathcal{L}_1^{\dagger}\Lambda_0^{\dagger}(\tau)\mathcal{L}_0^{\dagger}\ln\pi_0}d\tau +\ave{\mathcal{L}_1^{\dagger}\ln\pi_0 }\right)
\\
&=& -k_B \ave{\mathcal{L}_1^{\dagger} \Lambda_0^{\dagger}(t)\ln \pi_0}.
\eeqar
In the second equality, we used Eq.~(\ref{eq:rho1}), and in the last equality we used $\Lambda_0^{\dagger}(\tau)\mathcal{L}_0^{\dagger}\ln\pi_0 = \frac{\partial}{\partial \tau}\Lambda_0^{\dagger}(\tau)\ln\pi_0$ and $\Lambda_0^{\dagger}(0)=1$.
Assuming the existence of a new new stationary state, one can note that for $t\rightarrow\infty$ the derivative $\frac{d}{dt}S_1=0$. To see this we assume $\rho_1(\infty)=\pi_1$ such that $\mathcal{L}(\pi_0+\pi_1)=0$. Then, $\mathcal{L}_0\pi_1=-\mathcal{L}_1\pi_0$, which implies that for long enough times the first line in Eq.~(\ref{eq:dS1dt_app}) vanishes.

The first order correction to the entropy production $\sigma_1(t)$ is now given by  
\beqar
\sigma_1(t)&=&\frac{d}{dt}S_1 -\sum_j\frac{\mathcal{J}_1^\alpha}{T_\alpha}=
\\ \nonumber
&=&-k_B\sum_\alpha \tr{}{\mathcal{L}^\alpha_0(\rho_1(t))\ln\pi_0}+\tr{}{\mathcal{L}^\alpha_1(\pi_0)\ln\pi_0}
\\ \nonumber
&+& k_B\sum_\alpha \tr{}{\mathcal{L}_0^\alpha(\rho_1(t)) \ln \pi_0^\alpha } +\tr{}{\mathcal{L}_1^{\alpha}(\pi_0)\ln \pi_0^\alpha} 
\\ \nonumber
&+& \tr{}{\mathcal{L}_0^{\alpha}(\pi_0) \partial_{\delta}\ln\pi^\alpha|_{\delta=0}} 
\\ \nonumber
&=&-k_B\sum_\alpha \int_0^t\ave{\mathcal{L}\dg_1\Lambda\dg_0(\tau)\mathcal{L}_0^{\alpha\dagger}(\ln\pi_0-\ln\pi_0^\alpha)}d\tau
\\ \nonumber
&+&\ave{\mathcal{L}_1^{\alpha\dagger}(\ln\pi_0-\ln\pi_0^\alpha)}-\beta_\alpha\ave{\mathcal{L}_0^{\alpha\dagger}V}
\eeqar
It's worth noticing that at equilibrium (single bath),  the first order of the entropy production $\sigma_1(t)=0$. The change in the von Neumann entropy equal to the entropy flow from the bath. For nonequlibrium system this is no longer the case.

\bibliographystyle{ieeetr}

\bibliography{citeamikam}

\end{document}